\documentclass{article}
\usepackage{amssymb}
\usepackage{amsmath}
\usepackage{graphicx}
\usepackage{color,soul}
\usepackage{rotating}
\usepackage{afterpage}
\usepackage{helvet}

\usepackage{setspace}
\usepackage[a4paper, total={6.5in, 9in}]{geometry}
\usepackage{fancyhdr}
\usepackage{lastpage}
\pdfoutput=1
\usepackage{hyperref}
\hypersetup{
  pdfinfo={
    Title={Quantitative phase and polarisation endoscopy for detection of early oesophageal tumourigenesis},
    Author={George S. D. Gordon},
    Subject={Endoscopy, medical imaging, optical fibres},
    Keywords={}
  }
}
\usepackage{pdfpages}
\begin{document}


# Quantitative phase and polarisation endoscopy for detection of early oesophageal tumourigenesis

George SD Gordon[1,2,3], James Joseph[2,3], Maria P. Alcolea[4], Travis Sawyer[2,3], Alexander J Macfaden[1], Calum Williams[1,2,3], Catherine RM Fitzpatrick[1,2,3], Philip H Jones[5], Massimiliano di Pietro[5], Rebecca C Fitzgerald[5], Timothy D Wilkinson[1]*, Sarah E Bohndiek[2,3]*

[1] Department of Engineering, University of Cambridge, UK

[2] Department of Physics, University of Cambridge, UK

[3] Cancer Research UK Cambridge Institute, University of Cambridge, UK

[4] Wellcome Trust-MRC Cambridge Stem Cell Institute, University of Cambridge, UK

[5] MRC Cancer Unit, University of Cambridge, UK

* Corresponding author information:

Dr Sarah Bohndiek: *Address*: Department of Physics, Cavendish Laboratory, JJ Thomson Avenue, Cambridge, CB3 0HE, U.K. and Cancer Research UK Cambridge Institute, Li Ka Shing Centre, Robinson Way, Cambridge, CB2 0RE, U.K. *Phone*: +44 1223 337267; *Fax*: +44 1223 337000; *Email*: seb53@cam.ac.uk.

Prof. Tim Wilkinson: *Address*: Electrical Engineering Division, Department of Engineering, 9 JJ Thomson Avenue, Cambridge, CB3 0FA, U.K.; *Phone*: +44 1223 348353; *Email*: tdw13@cam.ac.uk.

Word count (Introduction, Results and Discussion): 2993
Figure count: 5

1

**Abstract**

Phase and polarisation of coherent light are highly perturbed by interaction with microstructural changes in pre-malignant tissue, holding promise for label-free early cancer detection in endoscopically accessible tissues such as the gastrointestinal tract. Flexible optical fibres used in conventional diagnostic endoscopy scramble phase and polarisation, restricting clinicians instead to low-contrast amplitude-only imaging. Here, we unscramble phase and polarisation images by exploiting the near-diagonal multi-core fibre (MCF) transmission matrix to create a novel parallelised fibre characterisation architecture, scalable to arbitrary MCFs without additional experimental overhead. Our flexible MCF holographic endoscope produces full-field en-face images of amplitude, quantitative phase and resolved polarimetric properties using a low-cost laser diode and camera. We demonstrate that recovered phase enables computational re-focusing at working distances up to 1mm over a field-of-view up to 750×750μm$^2$. Furthermore, we demonstrate that the spatial distribution of phase and polarisation information enables label-free visualisation of early tumours in oesophageal mouse tissue that are not identifiable using conventional amplitude-only information, a milestone towards future application for early cancer detection in endoscopy.

**Introduction**

White-light endoscopy is the standard-of-care for inspecting large areas of the gastrointestinal (GI) tract and lung for pre-malignant change (dysplasia) and cancer[1]. For example, Barrett's oesophagus is an acquired metaplastic condition that predisposes patients to the development of oesophageal adenocarcinoma. The cancer risk for Barrett's patients increases significantly in the presence of dysplasia, up to more than 30% per year[1]. Early identification of dysplasia enables curative intervention through simple endoscopic resection or radiofrequency ablation[2]. Unfortunately, the current surveillance procedure uses white-light endoscopy combined with random biopsy, which together show only a 40-64% sensitivity for dysplasia, leading to high miss rates[3]. The 5-year survival rate for oesophageal cancer is only 15%, yet can be as high as 80% when patients are diagnosed with early-stage disease[3], hence improvements in endoscopic early detection methodologies are urgently needed. While application of dyes can improve contrast[4] but their use lengthens procedure times and can lead to toxicities[5]; label-free approaches could better address the clinical unmet need for improved contrast of dyplastic tissue.

Localised proliferation of cells in dysplasia scatters light, creating abnormally distorted wavefronts[1]. Phase imaging has been shown to be highly sensitive to such distortions[6]. Furthermore, polarimetric imaging measurements of diattenuation, retardance and circularity can be modified by scattering, as well as perturbed by the higher concentrations of optically anisotropic molecules such as collagen abundant in tumours[7]. Light scattering spectroscopy has shown promise for detecting dysplasia using phase and polarisation information, but interrogates only a narrow field-of-view[8]; this limitation is partially mitigated by optical coherence tomography, which uses lateral scanning mechanisms and can be integrated into capsule endoscopes[9], but interpretation of the resulting cross-sectional images remains challenging[10]. In addition, both approaches require dedicated complex instrumentation, which has limited endoscopic application.

Flexible medical fibrescopes relay optical information from within the patient to the imaging system outside, which could be used to enable direct, wide-field, phase and polarisation imaging in existing clinically approved systems with comparatively simple and low-cost elements, such as coded apertures, gratings[11], and polarising optics[12]. Commercial endoscopes typically use distal sensors ('chip-on-tip') and although prototype devices with distal optics for other modalities have been developed (e.g. holographic imaging[13]) the additional bulk (>2-fold width) makes integration with existing endoscopic procedures difficult.

Fibre bundles are typically <1mm in width, independent of the imaging modality, making them attractive candidates for implementation of novel medical imaging technologies in endoscopy. Unfortunately, the optical fibre bundles used in flexible fibrescopes inherently scramble phase and polarisation information due to bending- and temperature-induced variations in glass refractive index, limiting diagnostic potential. Measured phase and polarisation could, however, be unscrambled by measuring and inverting the fibre transmission matrix (TM); a complex linear mapping between the two fibre facets[14].

Single-mode fibres (SMFs) with distal scanning[15], multi-mode fibres (MMFs)[16–18], and fused multi-core fibres (MCF) comprising many thousands of light-guiding cores[19] have been investigated for use as flexible endoscopes. SMFs only record point measurements so are unsuitable for recovering the relative phase between points on a wavefront without a high degree of laser and sample stability, though can produce polarimetric images through raster scanning[20,21]. While MMF TMs can in theory be predicted *a priori*[22], in reality, frequent *in situ* TM characterisation is necessary to account for thermal fluctuations and motion of the fibre, which represents a significant challenge. Several designs use phase and polarisation

information to recover MMF amplitude-only images[17], but cannot quantitatively determine phase or polarisation images. Furthermore, SMFs and MMFs have yet to be adopted for widespread clinical use.

In contrast, MCFs have a near-diagonal TM resulting in minimal amplitude scrambling, hence are already clinically approved in a range of indications[23]. MCFs have been used to record interferometric images[24]. However, the highly scrambling nature of MCF has limited such methods to imaging differences in phase between a reference image and an image of mechanically perturbed tissue. This significantly reduces the practicality of wide-area mapping of tissue for clinical deployment, which is best achieved using full-field holographic imaging via TM characterisation.

Previous investigations using TM characterisation in conventional non-sparse Hadamard[25] and Fourier bases[26] in MCFs have been limited to wavefront shaping and wide-field imaging of a single polarisation. By contrast, here we exploit the near-diagonality of the MCF TM to enable a novel parallelised *in situ* fibre characterisation that could be applied during endoscopic imaging and is scalable to higher resolution MCFs without additional experimental overhead. Our flexible holographic endoscope is, to the best of our knowledge, the first to produce full-field images of amplitude, quantitative phase and resolved polarimetric properties through an MCF. Relying on phase retrieval, it operates simply with a low-cost laser diode and one camera. We first demonstrate quantitative phase recovery enabling computational refocusing[16] and then analyse the spatial distribution of phase and polarisation to provide label-free visualisation of dysplasia within healthy oesophageal tissue taken from a mouse model of early disease. This represents an important first step towards future endoscopic applications that could enable earlier detection of cancer in endoscopic surveillance and screening programmes in the gastrointestinal tract and lung for improved clinical outcomes.

**Results**

*Holographic endoscopy*

To retrieve amplitude, quantitative phase and resolved polarimetric properties of a sample, we created the optical set-up shown in Figure 1a (see Methods). The sample is illuminated using an 852 nm laser diode, a wavelength in the 'optical window'[27] of minimal tissue absorption and fluorescence. A square subsection of the MCF field-of-view (200×200 µm$^2$) is used for imaging, performed in transmission mode, and the remaining cores carry a stable phase reference (see Supplementary Fig. 1, Supplementary Note 1).

The operation algorithm of the device is shown in Figure 1b. First, the translation, rotation and scaling of the proximal facet as it appears on the camera (relative to SLM1) is determined. This information enables reconstruction of the different calibration patterns projected onto the distal facet of the MCF using pre-computed phase holograms displayed on SLM1 (Methods, Supplementary Fig. 2). Then, by comparing these calibration patterns with measured outputs, the TM of the MCF is reconstructed. An identity matrix represents a perfect image-preserving fibre, while variation of diagonal elements and the presence of non-zero off-diagonal elements represent varying degrees of scrambling. Exploiting the near-diagonal structure of the TM, we are able to physically parallelise the characterisation process by scanning an array of 12 equispaced spots (spacing = 68.2 µm) across the MCF. The result is a 12-fold increase in speed compared to dense TM characterisation at the same resolution using a MMF (Methods, Supplementary Fig. 1, Supplementary Note 1). Finally, the inverse of the determined TM is multiplied by the raw complex image to reconstruct the amplitude, phase and polarisation of a sample placed at the distal facet (Fig. 1c). Imaging is then repeated with different illumination polarisation states to resolve polarimetric properties (Methods, Supplementary Note 2). By imaging known flat samples (glass slide for phase, quarter waveplate for polarisation), the phase error is estimated to be 0.3 rad and the polarisation retardance error 0.2 rad. The total acquisition time for an amplitude, phase and polarisation image set is 8.3 s, and the time taken to fully characterise the fibre is 50.8 mins.

*Figure 1. Overview of holographic endoscope design and operation for recovery of optical phase and polarisation.* a) Transmission mode operation of the holographic endoscope. Liquid-crystal spatial light modulators (SLMs) are combined with polarising beam splitters and waveplates to perform polarisation-diverse digital holographic imaging through a MCF. SLM1 illuminates the sample with the desired optical field and enables fibre characterisation, while SLM2 enables imaging of amplitude, phase and polarisation of light emerging from the fibre bundle. b) Overview of endoscope operation. c) Recovery of the amplitude, phase and polarisation from a test pattern displayed on SLM1 through the MCF by application of the determined transmission matrix. The text shape and properties input into the distal facet are accurately recovered from the measured image of the proximal facet. For comparison, the original distal image shown is sub-sampled at the same coordinates used for image reconstruction to represent the best possible recovered image. Bubble area indicates the power recorded within a given point. Image scale bar: 50 μm. Abbreviations: spatial light modulator, SLM; polarising beam splitter, PBS; half wave plate, HWP; lens, L; mirror M; arbitrary units, a.u.

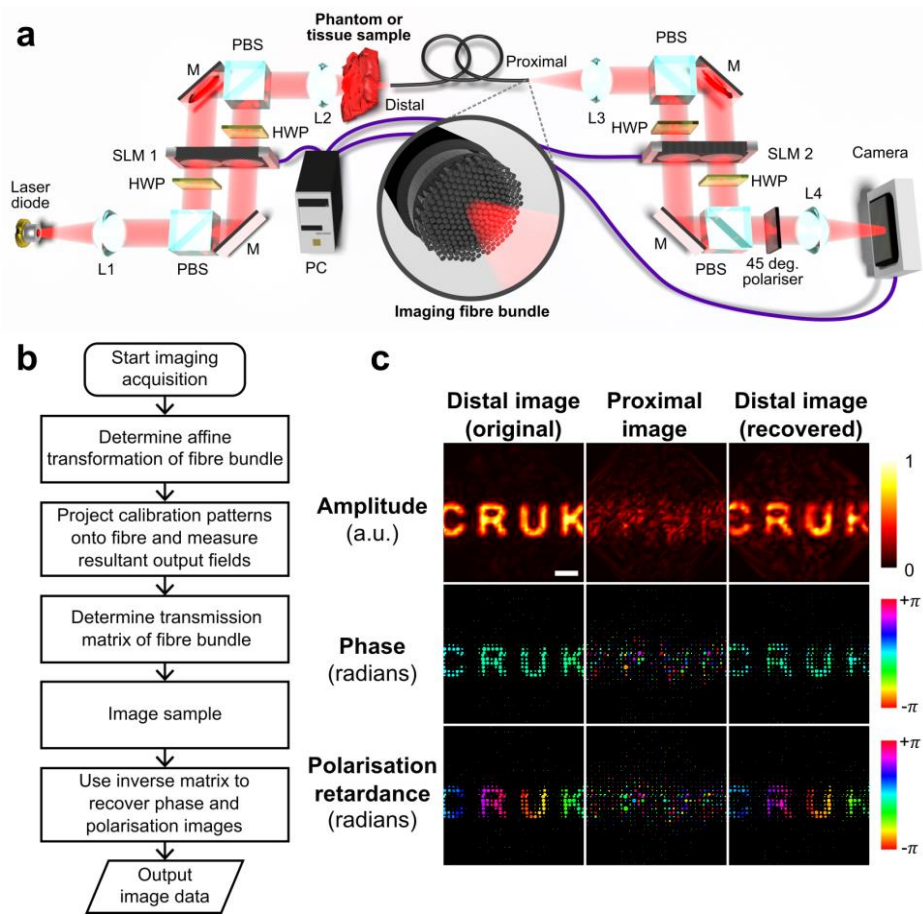

*Quantitative phase imaging*

We first evaluated recovery of quantitative phase information using a USAF 1951 resolution test chart placed at a range of working distances (Methods). Wide-area images of targets larger than the MCF image area (200×200 µm at 0 mm working distance) are formed by translating the sample and stitching the sub-images together. Analysis of the modulation transfer function using amplitude-only images of USAF chart line groups (Methods) shows a lateral resolution of 11.0±2.6 µm without TM correction and 9.0±2.6 µm with correction at a working distance of 0 mm (Fig. 2a). This is limited by undersampling of the distal facet during TM characterisation: the 12-spot characterisation pattern is translated in 9 µm steps, which could be reduced to 4.5 µm (the approximate core spacing). In practice, 9 µm steps reduced the dataset size for faster computation.

Next, the target is moved away from the distal facet in steps of 0.25 mm up to 1 mm and the quantitative phase information is utilised to correct defocus of the sample and improve resolution (Fig. 2b, Methods). The resolution halves to 18.0±2.6 µm in the uncorrected case, but returns to 10.0±2.6 µm with defocus correction applied (Fig. 2b, c). Given the measured fibre NA of ~0.35, the image area at 1 mm working distance is ~750x750 µm. Variable working distance provides a significant advantage over contact-based endoscopic techniques such as confocal laser endomicroscopy[3] and demonstrates that quantitative phase imaging allows a dynamically reconfigurable resolution and field-of-view, which in practical terms could enable wide area survey of the tissue, followed by a 'zoomed' analysis at high resolution for suspicious lesions.

***Figure 2. Holographic endoscopy is capable of phase retrieval and computational refocussing.*** *a) Resolution determination with a standard USAF 1951 test chart at 0 mm working distance is 11.0 ± 2.6 µm for the proximal facet image and 9.0 ± 2.6 µm for the recovered distal facet image. b) As expected, resolution inherently decreases with working distance due to defocus, but using phase information this can be corrected over a range of working distances (up to 1.0mm). c) Defocus correction maintains optimal resolution throughout the range of working distances tested (raw amplitude images: y = (10.5±2.7)x+(8.5±1.6), $r^2$ = 0.83, p = 0.031; focus corrected images, trend not significant, p = 0.54). Image scale bars: 200 µm.*

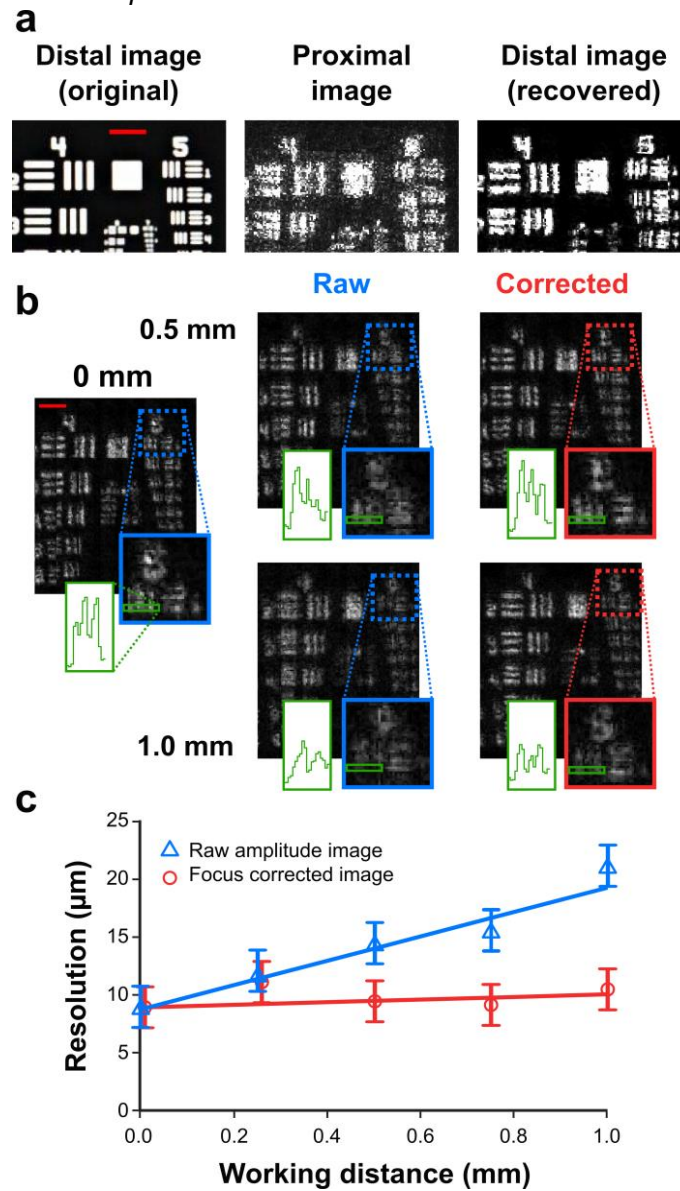

*Polarimetric imaging*

We then implemented a Jones calculus formalism to extract polarimetric properties of samples, an appropriate choice here because the spatio-temporal optical coherence makes depolarisation negligible; it also allows us to quantify relative phase between pixels[28] (see Methods). Each measured point on the sample is modelled as an elliptical retarder followed by a partial linear polariser (Supplementary Note 2.1, Supplementary Fig. 3), which gives 5 output model parameters: retardance axis orientation ($θ_φ$), diattenuation axis orientation ($θ_D$), diattenuation (*D*), retarder circularity (*δ*), and retardance (*φ*). The ordering of the retarder and polariser is chosen arbitrarily and if reversed will produce different results due to the non-commutativity of Jones matrices. However, if measured values of D are close to zero (standard deviation < 0.01), which they are for tissue samples measured here (standard deviation = 0.0086) then changing the ordering incurs an error of <5% (Supplementary Note 2.2, Supplementary Fig. 4). Bayesian inference is applied to pairs of input/output polarisation states to estimate these parameters (Methods, Supplementary Note 2.3).

Two polarisation test targets were used to validate polarimetric imaging: a birefringent resolution target (Fig. 3a), and a polarising resolution target fabricated in-house[29] using nanostructured metallic wire-grid polarisers (Fig. 3b). The birefringent target creates contrast via a 45° difference between the retardance axis in the foreground and background, i.e. structure is encoded only in $θ_φ$. This is observed in experimental measurements (Fig. 3a), and there is good agreement between simulation and experiment in this and across all other properties. For the polarising test target, contrast is encoded in $θ_D$ and this is observed experimentally (Fig. 3c). The spatial resolution of $θ_φ$ and $θ_D$ images was determined to be 36.0 ± 10.4 µm. The reduction in resolution compared to phase imaging is a result of averaging (joint inference) between neighbouring pixels to remove misalignment artefacts (Supplementary Note 2.4). Error propagation from raw phase and amplitude data is used to compute errors of each polarimetric parameter: $ΔD$=0.14, $Δθ_D$=0.28 rad, $Δδ$=0.12 rad, $Δφ$=0.26 rad, $Δθ_φ$=0.13 rad.

***Figure 3. Holographic endoscopy enables polarisation imaging.***
*a) Birefringent target with test patterns encoded in the retardance optic axis orientation show clear signals in this polarisation parameter, with negligible impact on other parameters. b) A custom diattenuation target was fabricated in-house from wire-grid polarisers with test patterns encoded in the diattenuation axis orientation, as illustrated by microscope and SEM images. Arrows on microscopy images indicate polarisation directions of vertical (left) and horizontal (right). c) Test target features appear strongly in the diattenuation axis orientation, with negligible impact on the diattenuation itself. Image scale bar: 100 µm (unless stated). Abbreviations: scanning electron microscopy, SEM.*

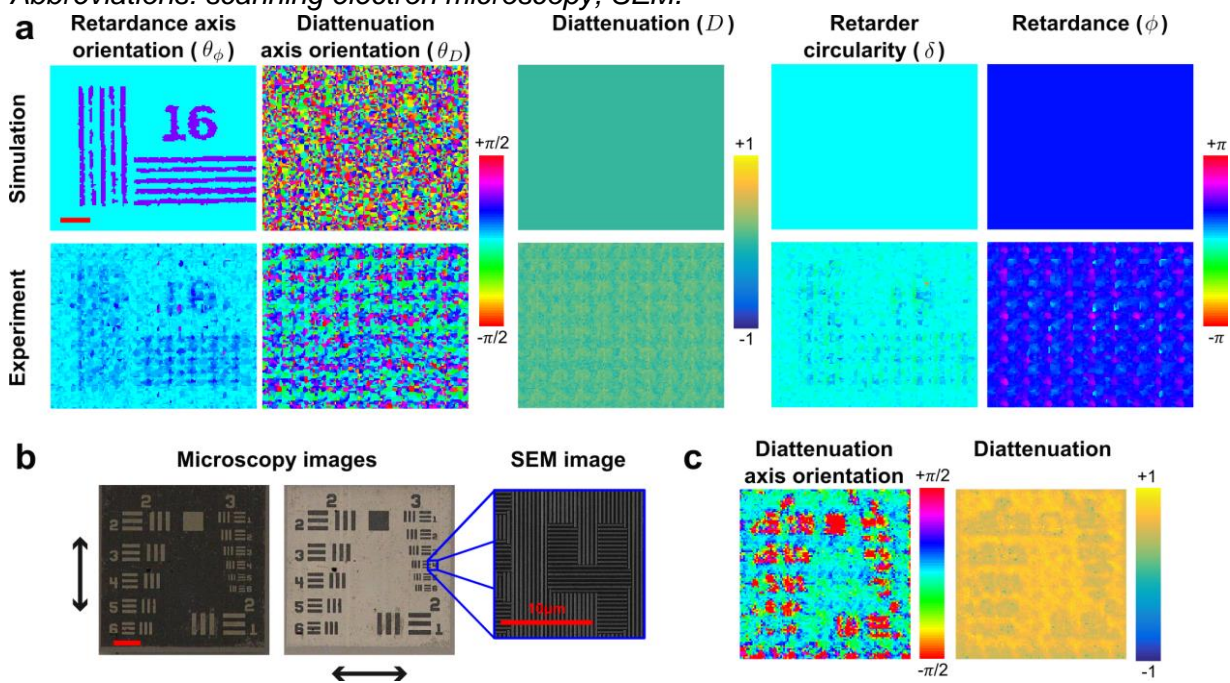

*Evaluation of biologically relevant variations in phase and polarisation*

Tissue-mimicking phantoms were used to emulate the expected scattering-induced phase variation and polarisation effects of diseased tissue. For scattering, we imaged phantoms made of polyacrylamide gel mixed with varying concentrations of intralipid (Methods, Supplementary Note 3). The raw phase images clearly show a more perturbed wavefront resulting from scattering (Fig. 4a). Spatial scattering-induced phase entropy (defined in Methods) increases significantly with intralipid scatterer concentration (Fig. 4a, b). For birefringence, we produced transparent phantoms using polyacrylamide gel, which has a strain-dependent birefringence (Methods, Supplementary Note 3). These samples were placed in a custom-designed stretching assembly and imaged at different strain levels (Supplementary Fig. 5). A clear change in retardance can be seen in the raw images and histograms (Fig. 4c), and a significant positive trend is observed as strain is increased (Fig. 4d, $r^2 = 0.71$). Other polarisation properties did not change significantly.

***Figure 4. Biologically relevant variations in phase and polarisation can be detected.***
*a) Raw phase images enable the scattering of light due to intralipid to be directly observed, while phase entropy images produce quantified scattering maps. b) Scattering-induced entropy shows a significant linear relationship [y = (0.20 ± 0.02) x + (0.25 ± 0.02), $r^2$ = 0.9816, p = 0.001; n = 2] with increased intralipid scattering concentration. Valid measurements are bounded by the effect of power-loss induced entropy, and for the last two samples this is the dominant effect and so they are excluded from the trend. c) Raw images showing recovered retardance, η, as a function of increasing strain with associated histograms. d) Retardance increases significantly as a function of strain [y = (2.91 ± 0.29) x + (0.06 ± 0.03), $r^2$ = 0.9725, p = 0.002; n = 4]. Image scale bars: 200 μm.*

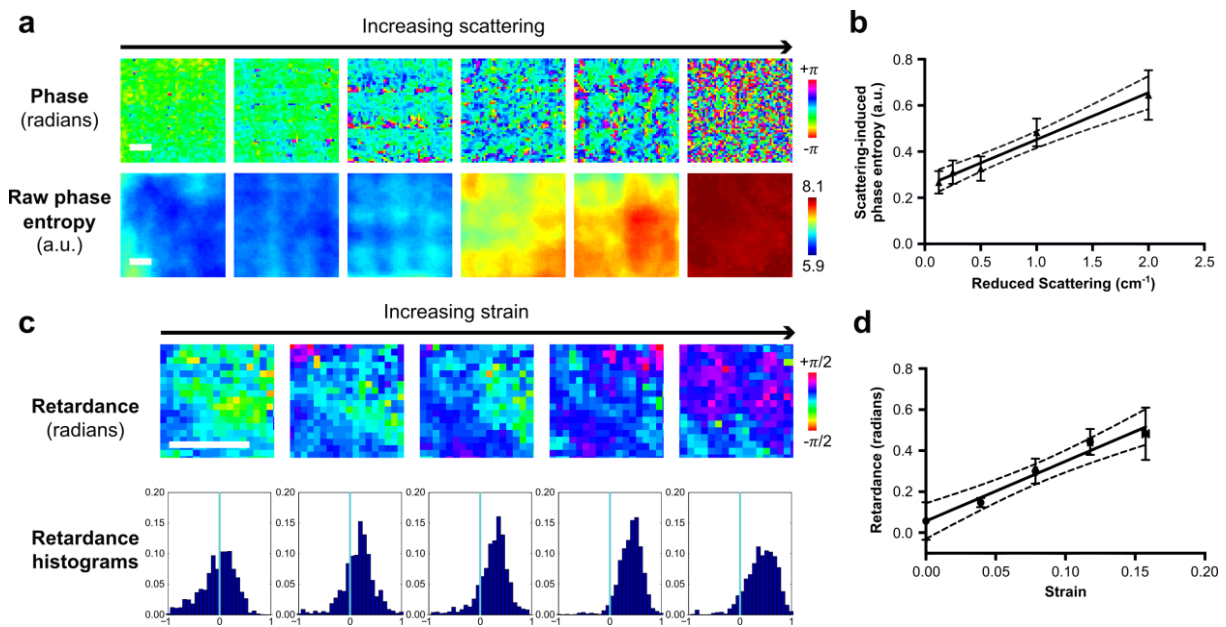

*Imaging early tumours in mouse oesophagus*

Finally, we sought to test the potential of our quantitative phase and polarisation resolved imaging to detect early lesions of the oesophagus, to pave the way for future application in early endoscopic detection of dysplasia in patients undergoing surveillance of Barrett's oesophagus. We imaged samples of mouse oesophagus containing abnormal tissue consisting of early pre-neoplastic tumours[30] (n = 6, 13 distinct lesions) and control tissues (n = 3, 7 healthy areas analysed, see Methods). The oesophagus is an ideal model for validating this technology because it is endoscopically accessible, exhibits well-defined multi-stage carcinogenesis that can be tracked, and has multiple stages of disease represented in space. Samples consisted of a thin layer of oesophageal epithelium, providing a representative model of surface scattering and polarimetric effects in the first few superficial cell layers where oesophageal tumours are usually detected by endoscopy. As for the phantom study, we used a spatial phase entropy filter, denoted H(Φ), to quantify scattering. The same filter is applied to the amplitude images to produce amplitude entropy, H(A), as well as to all polarimetric images. Additionally, a spatial averaging filter, denoted E(p), is applied to the raw amplitude, phase and polarimetic images so that the presented spatial resolution is the same as the entropy filtered images allowing comparison between the metrics (Supplementary Note 4). Note that only amplitude imaging is available in standard white-light endoscopy.

Figure 5a shows isolated sub-images of the measured samples over a selected range of modalities accessible with the holographic endoscope (Supplementary Fig. 6 shows an example whole-sample image set across all modalities). These are compared to fluorescence microscopy data, which is only available after applying a nuclear stain to the samples to highlight areas of abnormal tissue architecture and increased cell density routinely used to identify early tumours (Methods). While fluorescence microscopy methods have been applied *in vivo* in human patients to assist with biopsy targeting[4], their relatively narrow field of view prohibits application for wide-field endoscopic surveillance. Thus, we provide fluorescence microscopy data here as a gold-standard reference, rather than as an endoscopic comparator. Several label-free quantities measured with our holographic endoscope, particularly phase entropy, provide visibly higher signal in the lesion areas. Images taken using bright-field, phase contrast and crossed-polarised microscopy confirm that additional diagnostic information is encoded in phase and polarisation compared to amplitude alone (Supplementary Fig. 7). The non-zero diattenuation may arise from the assumption of negligible depolarisation (due to high spatial and temporal coherence, in contrast to previous models[31]), but has also been observed in breast and oral tumour tissue[32].

Contrast to noise ratio (CNR, defined in Methods) was computed separately for each of the 6 abnormal tissue samples to compare the diagnostic ability of the novel imaging modalities to identify lesions within a given tissue sample (Fig. 5b, full set of modalities shown in Supplementary Fig. 8). Phase entropy was found to provide significantly improved CNR compared to amplitude entropy (p = 0.0014), being comparable to the CNR obtained using the fluorescence nuclear stain (p = 0.150) but in this case acquired in a label-free manner. Individual polarimetric properties do not show significant increases over amplitude entropy, but when properties that share a common biological origin[33] – $\theta_D$, $\theta_\varphi$ and $\varphi$ – are summed, the resultant metric is significantly better than amplitude entropy (p = 0.034) and again comparable to fluorescence imaging (p = 0.160). It should be noted that the high variance observed within these CNR calculations is caused by inherent biological variation, as CNR was calculated separately for each sample, rather than technical limitation.

Finally, to demonstrate potential clinical application of the technique, we applied a binary classifier to produce a receiver operating characteristic (ROC) curve, showing the

sensitivity and specificity across the sample set (Fig. 5c). While the sample size is limited, encouragingly the phase entropy metric uniquely accessible with our endoscope performs significantly better than amplitude entropy (similar to the existing standard-of-care) and lies close to the fluorescence reference (available only in specialist centres), indicating superior performance for potential diagnostic use for this label-free metric compared to conventional amplitude-only imaging or fluorescence-based approaches that require the application of dyes. Phase entropy also provides the potential for very high specificity (>95% for 50% sensitivity), which could be advantageous for identifying low-risk cases (i.e. performing risk stratification) in surveillance programmes. The combined polarisation metric performs worse with binary classification than amplitude entropy, most likely due to the high inter-sample variance observed (Fig. 5b). The statistically significant CNR suggests there may, nonetheless, be valuable information contained in the polarisation data, but a more advanced classifier trained on a larger sample size would likely be required to evaluate the full potential of these data.

*Figure 5: Holographic endoscopy enables label-free identification of early lesions in the mouse oesophagus.* a) Composite image showing sections of healthy tissue and lesions from all 9 samples and 5 endoscope modalities. A range of focal and diffuse early tumour lesions can be observed in the reference fluorescence images. b) Contrast-to-noise ratio for the different modalities calculated independently for each of the 6 samples containing lesions (exact areas used are defined in Supplementary Fig. 6). The phase entropy and sum of $\varphi$, $\theta_\varphi$ and $\theta_D$ polarisation parameters produce contrast statistically comparable to fluorescence imaging, the gold standard reference modality ($p=0.148$ and $p=0.160$ respectively). Further, these two metrics produce significantly better contrast than amplitude entropy, the best available using conventional endoscopes ($p=0.0014$ and $p=0.034$ respectively). Scale bar 400 μm. Significance determined by paired two-tailed t-test,* $p < 0.05$, ** $p < 0.01$. c) Receiver operating characteristic curve illustrating performance of different modalities when a binary classifier with varying threshold is applied to discriminate between healthy and lesion tissue. Phase entropy significantly out-performs conventional amplitude imaging, but the large inter-sample variance of the combined polarisation metric results in reduced performance with a simple binary threshold classifier.

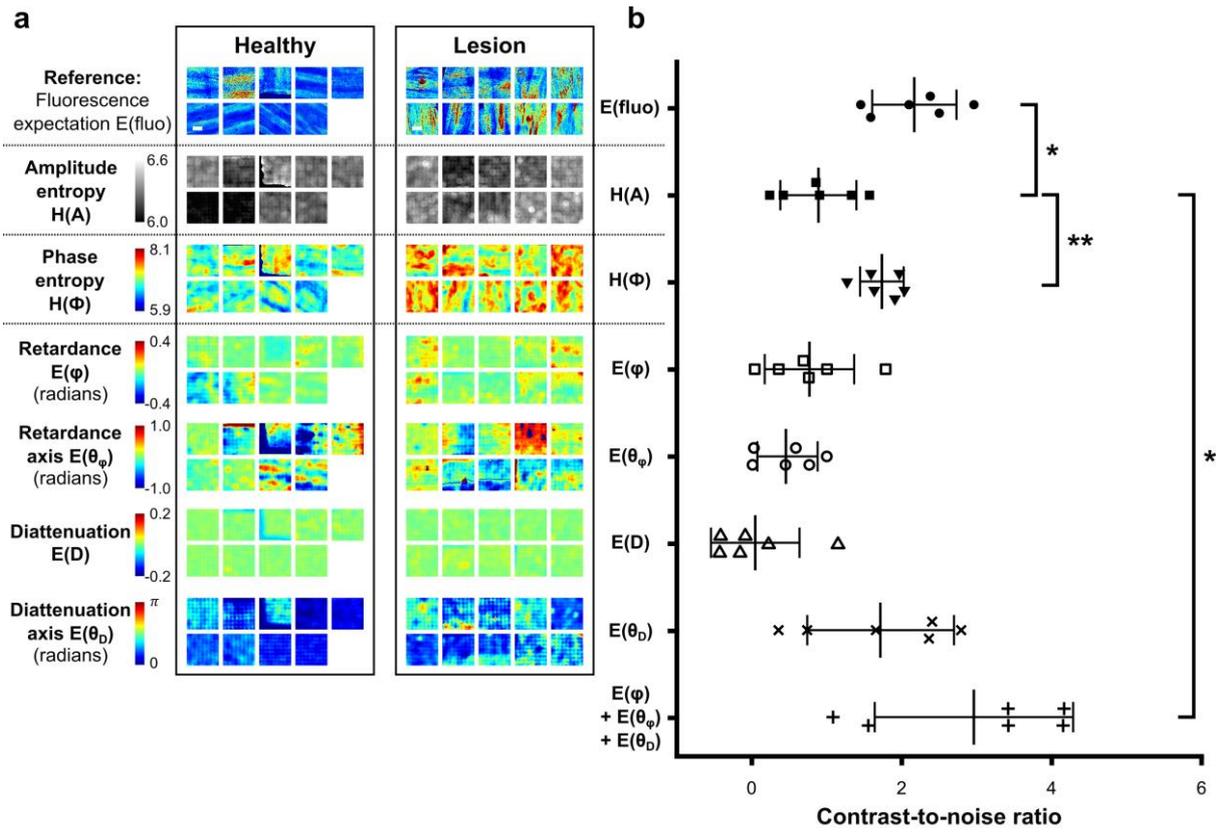
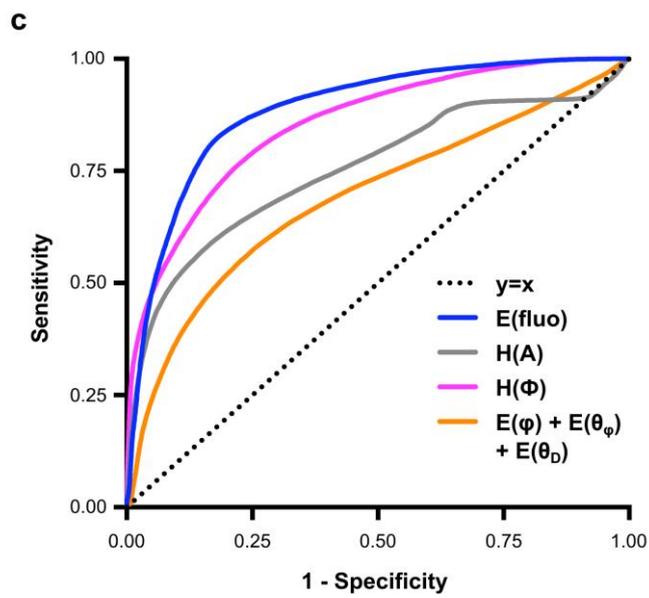

**Discussion**

Our holographic approach to parallelised TM characterisation enables endoscopic amplitude, quantitative phase and resolved polarimetric imaging through a flexible MCF for the first time. Flexible medical fibrescopes based on MCFs are already widely used in the clinic to relay optical information from within the patient to the imaging system outside. While some previous work has indirectly used phase and/or polarisation information to recover amplitude-only images from MMFs[17], our holographic endoscope is the first to directly retrieve wide-field en-face images of these properties together with amplitude information. By creating an architecture based around MCFs, our approach could be used with further developments to enable direct, wide-field, phase and polarisation imaging in existing clinically approved systems.

Our holographic endoscope records high resolution (9.0 ± 2.6 μm amplitude, phase; 36.0 ± 10.4 μm polarimetric) images at working distances up to 1 mm with field-of-view up to 750×750 μm. This limit arises from experimental space constraints but could in principle be extended to working distances of 1-2 cm and fields-of-view of ~1×1 cm, in line with conventional endoscopy. The ability to adaptively change working distance without distal optics is the first key strength of our approach. From the perspective of diagnostic endoscopy, it offers potential 'red-flag' surveying of large areas for abnormalities, followed by zooming in to perform high resolution 'optical biopsy'[34], which is currently impossible in a single device.

Quantitative phase is measured with an error of 0.3 rad while polarimetric properties are inferred with errors across the 5 parameters of <10%. The precision and resolution obtained in the quantitative phase and resolved polarimetric properties are the second key strength of our approach, being in principle sufficient to observe small, localised proliferations of cells associated with early tumorigenesis.

Quantitative phase and resolved polarimetric properties have shown promise for enhancing cancer detection in *ex vivo* and *in vivo* rigid endoscope studies[3,35]. Further, they are largely robust to variations in absolute intensity under varying measurement conditions. Retrieval of these properties through a flexible endoscope could therefore enhance contrast for pre-malignant and malignant changes during diagnostic endoscopy in the gastrointestinal tract and lung. To validate this in our holographic endoscope, we examined mouse oesophageal tissue containing early abnormal lesions and compared our findings to healthy control tissue. We found that the spatial entropy of phase information, a result of surface scattering, provided significant contrast improvement relative to amplitude-only images and was comparable to fluorescent images typically used to identify these early lesions *ex vivo*. Similarly, a simple sum of three measured polarimetric parameters ($\theta_D$, $\theta_\varphi$ and $\varphi$) provided significant contrast improvement compared to amplitude-only images and was comparable to the reference. Obtaining a label-free diagnosis equivalent to that provided by fluorescence staining avoids the need for extended procedure time and potential toxicities associated with *in vivo* dye application[5]. Importantly, these results indicate that phase and polarimetric images may contain more relevant diagnostic information than amplitude-only images, demonstrating a key advance over what is possible using current commercial endoscopes.

In contrast to previous work using MMF, the known near-diagonal structure of the MCF TM was used here to implement a novel parallelised architecture that offers a 12-fold speed increase compared to characterisation of a MMF with a dense TM. The reported parallelised architecture is scalable such that no extra experimental time is required for higher resolution MCFs with a greater number of cores. This is especially important when considering that, in real usage, the fibre will be bent or otherwise perturbed, requiring regular re-characterisation of the TM. An equivalent scheme could be implemented with MMFs by changing the optical

field representation basis from the point-wise basis used here to a basis matching the spatial eigenmodes of the fibre e.g. Laguerre-Gauss modes. However, the strong axial symmetry of such bases necessitates very precise fibre alignment, reducing robustness to movement and temperature changes typically encountered in clinical environments. Further, large pre-computed hologram libraries are required to create these modes[36], whereas the modes used here are generated ad-hoc by applying a phase tilt to a single precomputed hologram. A key advantage of our approach is that the point-wise basis does not require alignment with an axis and inherently benefits from a near-diagonal TM because of the pixellated structure of the MCF.

Although these results are promising, several key limitations to clinical translation remain, which are addressed in further detail in the Supplementary Information (Supplementary Note 5). Firstly, the current transmission-mode architecture must be adapted for reflection-mode imaging. Oblique back-scattering illumination can simulate transillumination in reflection[37] but *in situ* reflection-mode TM characterisation is a significant challenge that is only in the early stages of exploration. Theoretical work by Gu *et al* has proposed using light back-reflected from a known distal plate to dynamically update a pre-recorded TM, but this requires a distal shutter and assumes a unitary TM[38], which is not typically the case for real fibres[36]. We have shown in simulation that a multi-layered reflector stack placed on the distal end of a fibre that enables TM recovery in reflection mode for non-unitary fibre TMs and without a distal shutter [Patent filed]. Semi-reflective plates with the required heterogeneous polarisation properties for this application could be easily produced using standard nanofabrication techniques[29]. Further, experimental findings using a highly-spaced MCF for two-photon imaging are also encouraging[39]. These are promising indicators that experimental implementation of reflection-mode characterisation is feasible. It should also be noted that the need for reflection-mode characterisation is avoided entirely in applications where rigid endoscopes are suitable, for example, deep tissue imaging through a needle[35,40]. A further challenge of reflection-mode imaging is the collection of multiply scattered photons from deep within the tissue, which could potentially be overcome through the addition of methodology for depth sectioning. Using holographic imaging as presented here, some degree of depth sectioning may be obtained by adaptation of previously reported digital confocal imaging[41] or multi-wavelength approaches[42].

A second limitation is the rate of operation. Current speeds for amplitude, phase and polarisation imaging (8.3 seconds per image), polarimetric imaging (3×8.3=24.9 seconds), and TM characterisation (50.8 minutes) are limited by camera and SLM frame rates. *In vivo* motion artefacts from peristalsis or respiration typically demand imaging times of <0.05 s[43]. We estimate that it would be possible to reduce imaging time to <0.05 s for real-time *in vivo* operation and characterisation time to <1 min, using strategies such as: advanced phase/polarisation retrieval methods requiring fewer amplitude images (2 vs. the current 7); high-speed imaging using a higher frame rate camera or photodiode-based compressive imaging; and replacing LCoS SLMs with fast digital micromirror devices that can achieve up to 100x increase in speed for wavefront shaping[44]. Should fibre characterisation measurements require stability over a longer time period, this could be achieved using commercially available endoscopic balloon stabilisers[45]. TM computation and image recovery could be reduced to <1 min and <0.1 s respectively by replacing the current iterative approach (>1000 Fourier transforms required) with transport-of-intensity equation methods (two Fourier transforms and one derivative required)[46]. Furthermore, frameworks that avoid the need for explicit TM reconstruction and directly reconstruct the image data have already been reported[47].

Thirdly, examination of a greater diversity of lesion stages and tissue types, particularly from *ex vivo* human tissue, is required to further quantify the robustness of phase and polarimetric imaging for detecting early tumours in the oesophagus. While our simple binary classifier showed promising performance in our ROC analysis, with a larger dataset we could train a more advanced classifier that optimally combines all measured properties into a single contrast metric and enables disease staging.

In conclusion, we present the first flexible MCF to produce full-field en-face images of amplitude, quantitative phase and resolved polarimetric properties. Using this proof-of-concept holographic endoscope, we demonstrate that these additional optical properties enable dynamic refocusing at working distances up to 1 mm. Importantly, we also highlight the potential of quantitative phase and resolved polarimetric properties for label-free identification of early tumours in the oesophagus. Future work addressing limitations of the present approach may enable deployment for diagnostic endoscopy in the gastrointestinal tract and lung, particularly in surveillance of patients with Barrett's oesophagus where early identification of dysplastic lesions is crucial and enables curative intervention. This could ultimately lead to an instrument that improves early detection of cancer, and thus improves survival rates from oesophageal cancer.

**Materials and methods**
*Optical set-up*

A schematic overview of the optical set-up configured to achieve dual-polarisation operation[36] is shown in Figure 1a. Illumination is provided by a laser diode operating at 852 nm with a power output of 35 mW (DBR852S, Thorlabs). This wavelength is optimal for our MCF (FIGH-06-350G, Fujikura; 2m length, 6000 cores, core diameter ~2.9 µm, core spacing 4.4 µm, outer diameter 350 ± 20 µm) because inter-core coupling is low enough to preserve the approximate image shape while still ensuring single-mode light guiding in each core. This not only enables relatively easy alignment, but is also ensures a near diagonal structure, which is critical for the parallelised TM characterisation method.

Spatial light modulators (SLMs; 1920 x 1080 resolution, PLUTO-NIR-015, Holoeye) are used for control of both illumination (SLM1) and detection (SLM2). One SLM characterises the fibre and illuminates the sample while the other performs phase shifting interferometry between polarisations for imaging. The laser diode output is split into two orthogonally polarised components using a polarising beam splitter (PBS102, Thorlabs) and the 2:1 aspect ratio of the SLM is then exploited, enabling each half of SLM1 to control one of the two orthogonal polarisations. The nematic liquid crystal on silicon (LCoS) SLM is used in phase-modulation mode to introduce a programmable phase shift (0 to 2π) at each pixel. This requires horizontally polarised incident light, and produces vertically polarised reflected light. Half wave plates (WPF-2-12.7x12.7-H-850, FOCtek Photonics) are therefore used in one arm to convert vertically polarised incident light to the required horizontal polarisation, and in the other arm to restore vertically polarised reflected light to its original horizontal polarisation. The two arms are recombined via a second PBS. Our illumination path design provides the ability to perform independent spatial modulation and phase shifting of polarisation states to create arbitrary polarisation wavefronts.

Samples are mounted in an automated stage (NanoMAX TS XYZ, Melles-Griot). Samples larger than the endoscope field of view (FOV) were translated in front of the distal facet of the MCF with a fixed ~15% overlap between positions, which enabled stitching and phase matching of adjacent images. Light passing through the sample is collected by the distal end of the MCF. The resulting FOV is approximately 200×200 µm at 0 mm working distance to the sample.

With no sample in place the total power coupled into the imaging MCF is ~0.6 mW. The lenses coupling light into the MCF are selected to enable addressing of large areas (compared to the core size) and so their numerical apertures are mismatched with the numerical apertures of individual cores. This mismatch, combined with inherent SLM inefficiency and the use of phase-only holograms accounts for the power loss observed between the laser and the MCF. Taking this into account, it is calculated that typical imaging operation uses 3.2 mWcm$^{-2}$ in the sample plane. For comparison, power densities of > 3000 mWcm$^{-2}$ have been FDA-approved for *in vivo* fluorescence imaging of the human oesophagus[48]. At the imaging output (proximal end) of the MCF, the beam is again split into two arms via a polarising beam splitter (PBS102, Thorlabs), with each half of SLM2 modulating each arm. These arms are recombined and passed through a linear polariser oriented at 45° to the axes of the polarised components. Our detection path design allows phase-shifting interferometry between the two polarisation states in order to determine their phase relationship.

The light is finally detected at the camera (4.2 megapixel, 90 fps MQ042RG-CM, Ximea) as an amplitude-only image. The camera is operated in 8-bit mode to achieve the maximum frame rate. To accommodate the different illumination powers experienced during

TM determination (no sample present) and sample imaging, images are captured at several exposure times between the maximum (40 ms) and minimum (20 μs) and a composite high dynamic range image is composed (Supplementary Fig. 9). The SLMs and camera are synchronised via a GPU accelerated PC and controlled using custom software written in C/C++. Unlike with multimode fibres, where the input field at the distal fibre facet must be imaged with a second camera during calibration[17], using an imaging MCF preserves the approximate form of the images so the input field can be reconstructed from a single camera using accurate simulations of the holograms displayed on the SLM (see Supplementary Note 1). The footprint of the entire system is 300 mm x 600 mm, built on an optical breadboard.

*Endoscope operation*

The overview of the endoscope operation procedure is shown in Figure 1b. Before generating holograms for TM determination and sample illumination, two calibration steps must be performed. First, aberration correction is implemented to correct for curvature of the SLM surface. This is achieved by maximising the peak height of a focused spot created by a single-mode fibre imaged onto the camera combined with several neutral density filters (NEK01, ThorLabs) and recorded by the camera. A hybrid gradient-descent/genetic algorithm was used to find the optimum weighting of a limited number of Zernike polynomial 2D surface correction terms. By performing this optimisation *in situ* we also compensate for aberrations introduced by the other optical elements, and ensure that the amplitude and phase of the optical fields used during fibre characterisation are accurate, later enabling us to simulate these during TM estimation (see Supplementary Note 1).

The second calibration step requires the translation, rotation and scaling of the image of the MCF as it appears on the camera to be assessed relative to the actual orientation of the distal facet of the MCF. To achieve this, a single spot is raster scanned over the imaging area on the distal facet by displaying a blazed grating on SLM1. The centroid of each spot position on the camera is found. The fitted centroids are then related through an affine transformation to known input spot locations on the distal facet. With these calibrations complete, it is possible to reconstruct the optical field input to the distal end of the MCF using a holographic approach.

Three distinct holograms are used for operation of the endoscope. The first is an array of 12 equispaced spots (Supplementary Fig. 2a) used for characterisation of the MCF TM before the introduction of the sample. The TM ($A^{-1}$) describes the relationship between the optical fields at the distal and proximal facets of the MCF and must be determined in order to retrieve amplitude, phase and polarisation images from our system. The choice of 12 spots ensures individual spots are spaced sufficiently far from each other that there is negligible interference between them at the proximal facet of the MCF. Using an array of spots physically parallelises the characterisation process (compared to the conventional approach of raster scanning a plane wave[16] or scanning through a full Hadamard basis[14]), enabling 12-fold faster determination.

To compute the hologram, the illumination profile incident on the SLM – a collimated laser beam – is multiplied by a pixellated phase profile to be displayed on the SLM and the Fourier transform is taken to establish the image expected at the focus of the second lens in the 'replay field' (Supplementary Fig. 2a). The pixellated phase profile, or hologram, is then optimised to produce the desired replay field. As the hologram is phase-only, light can only be redistributed, not attenuated, which limits the range of possible replay fields for a given illumination. Therefore, to enable generation of arbitrary patterns for a given illumination, a region of interest (ROI) is defined to restrict the part of the replay field that is constrained during the optimisation. Optimisation is then achieved by minimising the difference between

21

the ROI in the replay field and the target field using a simulated annealing algorithm to find a globally optimal hologram.

The camera used is inherently sensitive only to intensity (the square of amplitude) and so to reconstruct images of phase and polarisation, multiple intensity-only frames must be recorded. The square root of these provides amplitude images, which are then input into an iterative phase-retrieval algorithm. This simplifies the experimental system because, unlike interferometric designs, it does not have the requirement for high stability and works with a comparatively low coherence (~4m coherence length), compact and low-cost laser diode illumination source. Interferometric designs typically require expensive Ti:Sapphire or fibre lasers[17] and can suffer problems with drifting of the phase reference.

In total, 11 frames are required to recover a full amplitude, phase and polarisation image. To infer polarimetric properties of a sample, a second and (optional) third set of 11 frames are acquired using different illumination polarisation states. This process begins by establishing the phase of the horizontally polarised image component. The vertically polarised arm of the endoscope is disabled by displaying a chequerboard pattern on the corresponding half of SLM2, which scatters most of the light outside the camera aperture. A stack of amplitude-only images from different focal planes is then recorded by displaying a series of different Fresnel lenses (parabolic phase masks) on SLM2[46] in the horizontally polarised arm. Since SLM2 is located in the Fourier plane, light propagation between each set of adjacent focal planes can be simulated to enable phase retrieval. We simulate light propagation between adjacent focal planes, as determined by the different parabolic phase masks (Zernike $Z_2^0$) on SLM2, replacing the amplitude information at each with the recorded image while keeping the phase constant; this approach rapidly converges on an appropriate phase profile and is known to be robust in the presence of noise[46]. To further improve convergence, the phase was partially randomised at the completion of each iteration, with the level of randomisation reducing over time, akin to the cooling curve in simulated annealing.

The problem of phase reconstruction of an image stack is non-convex and thus convergence of the iterative approach is not guaranteed[46]. Therefore, the focal offsets and number of planes must be empirically adjusted until reliable convergence in suitable timescales is achieved. Here, the focal offset is constrained to be relatively large in order that light from opposite sides of the fibre facet can fully interfere. This allows for phase imaging of objects with large areas of low or no light (e.g. 1951 USAF resolution test target). Under this constraint it was found empirically that seven focal planes was the minimum number that produced reliable convergence in an acceptable timeframe (<1 minute/250 iterations).

With the phase of the horizontally polarised image determined, the phase in the vertical polarisation can be determined by establishing the phase-offset between the two polarisations in each pixel. This is achieved by creating a phase shifting interferometer between the two polarisations. The 45° polariser between SLM2 and the camera results in coupling of the horizontal and vertical polarisations into a common basis, allowing interference. Using SLM2, a fixed phase delay is added to the horizontally polarised arm which is swept in increments of (n-1)/2π for n = 4 images. A sinusoid is then fitted to the intensity in the amplitude-only image at each pixel to determine the phase shift.

The mathematical framework and experimental methods for determining the MCF TM **A**$^{-1}$ are given in detail in Supplementary Note 1 and illustrated in Supplementary Figure 1. Experimentally, the hologram in Supplementary Figure 2a is displayed on SLM1 to create an array of 12 equispaced spots in one of 3 elliptical polarisation states, which are then scanned over a 9×9 grid. The choice of grid size ensures that about half the cores in either dimension inside the imaging area will be characterised. This reduces the TM size by a factor of 4, which

is necessary so it can be stored in computer memory for faster processing. A minimum of two polarisation states are required to resolve the two orthogonal polarisations, but three are chosen here as a trade-off between reducing noise and limiting experimental time. Elliptical states are used because they can be generated without blocking either of the two orthogonally polarised arms, resulting in maximum utilisation of available power and hence reduction of measurement noise. The recorded images are split into 12 sub-images, each containing one of the illuminating spots, producing 12×81×3 = 2916 images for matrix determination. Following correction for piston and phase tilt, the inverse problem is solved using a basis pursuit denoising algorithm implemented in the SPGL1 package[49] with an L1 regulariser. The L1 regulariser enables further noise reduction by favouring sparse solutions.

Once the TM $\mathbf{A}^{-1}$ has been determined, the second of the three holograms used for operation of the endoscope can be applied. The second hologram provides a test pattern that can be used to confirm the accuracy of the TM result. A short text phrase (CRUK) provides a useful pattern for this purpose (Supplementary Fig. 2b). The third hologram, used for sample imaging, should ideally provide a uniform illumination with a flat phase front. However, since this is passed through the same lens that is used to produce the array of spots, in reality the amplitude profile converges to a Gaussian profile while the phase profile is flat to within ±0.15 rad (Supplementary Fig. 2c). To achieve high fidelity within the replay field ROI, it is often necessary for the hologram to direct much of the light to regions outside of the ROI, meaning that the ratio of light power within the ROI to the total in the replay field may be as low as 1.7% during sample imaging.

*Image data recovery and analysis.*

All image processing was performed in MATLAB (Mathworks) unless otherwise stated. The resulting phase and polarisation images ($\mathbf{y}_{meas}$) are multiplied by the fibre TM $\mathbf{A}^{-1}$ ($\mathbf{x}_{dist} = \mathbf{A}^{-1} \mathbf{y}_{meas}$) in order to recover the optical field at the distal facet of the MCF ($\mathbf{x}_{dist}$, see Supplementary Note 1). Successful phase imaging makes it possible to correct for any defocus arising from the sample being moved away from the distal facet by simulating the propagation of the distal field through free-space using the Fresnel diffraction integral:

$$E(x,y,z) = \frac{e^{ikz}}{i\lambda z} \iint_{-\infty}^{+\infty} E(x',y',0) e^{\frac{ik}{2z}[(x-x')^2 + (y-y')^2]} dx' dy' \qquad (1)$$

where z is the distance between the desired focal plane and the distal facet, x and y are spatial coordinates in the desired focal plane, and x' and y' are spatial coordinates in the distal facet plane. Discrete Fourier transform methods were used to compute this integral efficiently following appropriate sampling as previously reported[50], and z is either input directly if the physical displacement is known or identified by inspection of the focus in the final image if physical displacement is unknown. Correction for this defocus enables resolution to reach the numerical aperture limited optimum for the given degree of undersampling at the distal facet.

**Polarisation properties.** A Jones matrix formalism is used here to model the polarimetric properties of samples. Jones calculus is applicable here because the light is temporally and spatially coherent due to the laser diode and the single-mode filtering of the cores respectively, thus depolarisation is negligible[28]. Further, Jones calculus can describe phase shifts between pixels, which is essential for quantifying scattering-induced phase variation. As detailed in Supplementary Note 2, Bayesian inference using a maximum likelihood optimiser (the STAN package[51]) was applied to each set of three amplitude, phase and polarisation images to extract the individual polarimetric sample properties of: diattenuation ($D$), also called linear dichroism; diattenuation axis ($\theta_D$); retarder circularity ($\delta$); retardance ($\varphi$); and retardance axis ($\theta_\varphi$).

**Entropy.** Aside from directly inferred values of phase and polarisation, the level of spatial scattering of these parameters is also informative. To quantify this, we compute the local entropy of a given property at each pixel (x,y), denoted H(g(x,y)), where g() may be an amplitude or phase value, or a polarimetric property. Entropy quantifies the spread of a data set or probability distribution making it a natural choice to quantify scattering. Consequently, it has previously been applied for cancer identification in amplitude only images[52]. To compute local entropy at point (x,y), we select a set of neighbouring pixels and fit a probability distribution to the pixel values. For amplitude data, a Gaussian distribution is used, but for circular quantities, such as phase and retardance, a von Mises distribution is used. The differential entropy of these fitted distributions is then computed to produce the final value assigned to position (x,y). This process is effectively a 2D spatial filter that quantifies how close to being uniformly distributed the values in the filter window are (further details in Supplementary Note 4 and Supplementary Fig. 10). The distribution fitting assumes independence of neighbouring pixels, but if this assumption is relaxed and instead a joint distribution between pairs of pixels a fixed distance apart is fitted (i.e. a bivariate distribution), the entropy is termed the *co-occurrence entropy*. Co-occurrence entropy is most often seen in the form of a grey-level co-occurrence matrix (GLCM) analysis, however, co-occurrence entropy was not found to differ significantly from standard entropy so was not included as a diagnostic metric in our evaluations. A full comparison between all computed metrics is given in Supplementary Figure 8.

These raw entropy metrics give a good indication of scattering-induced phase variation (Fig. 4a). However, they are also dependent on power because during the phase retrieval process, lower power pixels result in higher uncertainty in phase and therefore higher entropy. To isolate the specific influence of the scattering properties of the sample on entropy, we define scattering-induced entropy as:

$$H_{scat}(x,y) = [H(\beta_{dist}(x,y)) - H_{pow}(A^2(x,y))] / H_{pow}(A^2(x,y))$$

where $H(\ldots)$ is the entropy function, $\beta_{dist}(x,y)$ is the parameter of the distribution fitted to the phase of pixels neighbouring (x,y), $H_{pow}$ is the power-loss induced entropy determined experimentally using a series of neutral density filters, and $A^2(x,y)$ is the average power level for the neighbouring pixels of (x,y).

*Endoscope performance evaluation.*

Following the simple test pattern validation to confirm the accuracy of the TM, successful phase imaging was confirmed experimentally by performing computational refocusing. A standard 1951 USAF resolution test target (R1DS1N, ThorLabs) was placed at 0 mm working distance from the distal end of the imaging MCF, then displaced in 0.25 mm increments. At each displacement, the lines of known spacing on the target were used to compute the contrast as a function of spatial frequency, i.e. the modulation transfer function (MTF). Threshold values for the MTF used to determine resolution range from 0 to 0.5 in the literature[53]. We note for all 5 focal distances tested that MTF < 0.3 results in only 2 of the 3 lines being distinguishable by eye. We therefore selected MTF = 0.3 as a threshold. For each displaced target, linear interpolation is used to find the spatial frequency at which MTF = 0.3, which is used to define the image resolution and associated error.

Errors in the amplitude and phase measurements are determined by measuring a known flat sample (a glass slide) and observing the variance of the amplitude and phase over this surface. Such variance is caused by errors in the illumination due to the limitation of

hologram generation, and also due to noise in the system. The error in phase between polarisations is determined by imaging a known, flat waveplate (WPF-2-12.7x12.7-H-850, FOCtek Photonics) and observing the variance of this quantity over the image.

The decomposition of polarisation images into individual polarisation properties ($D$, $\theta_D$, $\delta$, $\varphi$, $\theta_\varphi$) was evaluated experimentally using two polarisation test targets. The birefringent target (R2L2S1B, Thorlabs) uses the National Bureau of Standards 1963A resolution pattern, with the lines encoded in a highly birefringent liquid crystal polymer. The foreground has the polymer retardance axis oriented at 0° and the background at 45°, testing our ability to recover the retardance property $\theta_\varphi$. To test the recovery of the diattenuation properties $D$ and $\theta_D$, we fabricated in-house a bespoke test target using electron beam lithography to produce metallic nanowire gratings with element width and pitch of less than 100 nm[29]. The resulting optical structure is a very efficient polariser (termed a wire-grid polariser) as the pitch is much smaller than the illumination wavelength. We encoded a standard 1951 USAF resolution test target with wire-grid polarisers in orthogonal orientations: the foreground is a polariser oriented at 0° and the background is a polariser oriented at 90°, as shown in Figure 3b. Using the lines encoded on these targets, the above approach for measuring resolution using MTF was again applied. The measured resolution for retardance axis and diattenuation axis were in agreement with that predicted when considering the spatial averaging performed on raw measurements during polarimetric inference in order to remove misalignment artefacts (Supplementary Note 2, Supplementary Fig. 10). Polarimetric properties are inferred solely from measured amplitude and phase in two polarisations, hence error propagation was used to determine errors in polarimetric properties by propagating errors in amplitude and phase through the Jones matrix model of an elliptical retarder and partial linear polariser.

*Imaging tissue-mimicking phantoms.*

The sensitivity of the holographic endoscope to biologically relevant concentrations of scattering and birefringence through phase and polarisation imaging was evaluated using tissue mimicking phantoms prepared as described in detail in Supplementary Note 3. Briefly, optical scattering phantoms (n = 2) were prepared by adding varying volumes of pre-warmed intralipid (20% emulsion, I141, Sigma-Aldrich) to liquid 1.5 w/v% agar solutions (05039, Fluka) to provide reduced scattering coefficients in the range from 0.125 to 2 cm$^{-1}$. 0.75% volume of 0.5 mgml$^{-1}$ nigrosin (198285, Sigma-Aldrich) was also added to provide an absorption coefficient of 0.005 mm$^{-1}$. Birefringent phantoms (n = 4) were prepared using Acrylamide/Bis-acrylamide, 30% solution (A3699, Sigma-Aldrich) catalysed using tetramethylethylenediamine (TEMED, T9281, Sigma-Aldrich) and ammonium persulfate (APS, A3678, Sigma-Aldrich). For both scattering and birefringent phantoms, the liquid solution was poured into petri dishes (CELLSTAR® 628-160, Greiner Bio-One) and then allowed to set at room temperature. For imaging, the birefringent phantoms were mounted on a custom fabricated mechanical stretcher (Supplementary Fig. 5) to create a strain-induced birefringence over the endoscope FOV. Optical properties of the phantoms were confirmed using a double integrating sphere system (reduced scattering and absorption coefficients) or polarisation state analyser (birefringence).

*Imaging biological samples.*

To test the ability of holographic endoscopy to identify tissue abnormalities in a label-free manner, we used *ex vivo* samples of mouse oesophagus from healthy untreated controls (n = 3; 7 healthy areas analysed) and carcinogen treated animals (n = 6; 13 distinct lesions analysed). Oesophageal tumours were induced by using a chemical carcinogen derived from

cigarette smoke, DEN (diethylnitrosamine, Sigma-Aldrich). The carcinogen was administered at a concentration of 0.04 ml l−1 in sweetened drinking water to induce sporadic mutations. Adult animals (over 10 weeks old) were treated three days a week for 8 weeks as previously described[30]. After carcinogen treatment, animals were aged from 6-9 months. Controls represent untreated animals. For epithelial tissue preparation, the oesophagus was cut open longitudinally, dissected into rectangular pieces of approximately 5 mm by 8 mm and incubated for 2–3 h in 5 mM EDTA at 37 °C. The epithelium was then carefully peeled away from underlying connective tissue with fine forceps in order to ensure that it was suitable for transmission mode imaging. Such preparation would not be required for practical application in reflection mode. The tissue was fixed in 4% paraformaldehyde in PBS for 30 min. For fluorescent nuclear staining, the tissue was incubated overnight with 1 µg ml−1 DAPI in PBS. For imaging, the oesophageal epithelium was then laid flat on a glass side and a coverslip was placed on top and sealed.

All experiments were approved by the University of Cambridge local ethical review committees and conducted according to Home Office project licenses P14FED054 and 70/8866. Experimental mice were doubly transgenic for the inducible Cre allele AhcreERT and the conditional reporter allele EYFP targeted to the Rosa 26 locus (AhcreERTR26flEYFP/wt)[30]. Strains were maintained in a C57Bl6 background. Experiments were carried out with male and female animals. No gender specific differences were observed.

To identify healthy regions and those containing early abnormalities or definite lesions, reference measurements were taken using bright-field, phase-contrast and polarisation microscopy (BX51-P, Olympus), as well as confocal fluorescence microscopy after staining with DAPI in a Leica TCS SP5 II system (optimal pinhole; speed 400 Hz; line average 3; resolution 1024×1024; image reconstruction using Volocity 6, PerkinElmer). Abnormal areas and lesions were identified on the confocal fluorescence microscopy images by an expert observer (M.A.) who has previously validated the fluorescence staining approach for identification of abnormal areas and lesions in this mouse model using genetic analyses[30]. These regions are shown on Supplementary Figure 7. A semi-automated registration process was then applied. The microscopy images were registered using the edges of the tissue samples as corresponding points between the samples; a 2×2 transformation matrix was then found and singular value decomposition used to obtain scale and rotation. A similar process was then applied to find the transformation between the phase-contrast microscopy image and the stitched endoscopic phase image, using striations in the tissue as corresponding points.

The mouse oesophagus samples are sufficiently thin that they have very high power transmission and the power-loss induced entropy is negligible so that $H_{scat}(x,y) \approx H(\beta_{dist}(x,y))/k$, where $k=H_{pow}(A^2(x,y))$ is a scale factor approximately constant across all samples and spatial positions. Therefore $H(\beta_{dist}(x,y))$ is used as the contrast metric for tissue samples. The potential utility of each of the phase and polarimetric properties that can be extracted using the holographic endoscopy was evaluated by calculating the contrast-to-noise ratio (CNR) between lesions and healthy tissue (regions defined in Supplementary Fig. 7) for each parameter, $p$, as:

$$CNR = \frac{\mu(F\ p_{lesion})-\mu(F(p_{healthy}))}{\sigma(F(p_{healthy}))} \qquad (2)$$

where F() represents either entropy, H(), or expectation, E(), depending on the parameter used, µ() represents the mean of a quantity and σ() represents the standard deviation of a quantity. The expectation operation, E(), serves the purpose of ensuring that spatial averaging of quantities is the same as for entropy, H(), so that the quantities are comparable. This

reflects the fact that smoothed raw values might represents useful diagnostic properties. For example an increase in retardance (φ) might be observed due to dense networks of collagen, while an increase in retardance angle entropy (H($\theta_\varphi$)) might be observed if the orientation of these collagen molecules is randomised.

CNR effectively represents a normalised test statistic that an area of lesion tissue was actually drawn from a healthy distribution. This normalisation enables comparing across modalities. For each CNR value computed, the lesion values are taken from the specific sample tested (aggregated across all lesions if there are multiple), but the healthy values are aggregated from across all 9 samples. Because it accounts for different noise levels, CNR can be used to compare the results obtained using fluorescence, amplitude, phase and polarisation measurements. This metric is computed for the DAPI images as a reference, then for the amplitude entropy, phase entropy, retardance expectation, retardance axis expectation, diattenuation expectation and diattenuation axis expectation and compared using a paired two-tailed t-test.

The receiver operating characteristic curve is computed for a quantity *Q* (e.g. fluorescence expectation, phase entropy) by applying a threshold, $t_Q$, to each pixel, *p*, in the data sets *L*, which contains all pixels situated in a lesion region across all samples, and *R*, which contains all pixels situated in a healthy region across all samples. This produces thresholded data sets defined as:

$$L_T\ p\ =\ \begin{cases} 0, & L\ p\ \leq t_Q \\ 1, & L\ p\ > t_Q \end{cases} \quad \text{and} \quad R_T\ p\ =\ \begin{cases} 1, & R\ p\ \leq t_Q \\ 0, & R\ p\ > t_Q \end{cases}$$

Sensitivity and specificity are then computed respectively as:

$$\frac{\sum_P L_T(p)}{|L_T|} \quad \text{and} \quad \frac{\sum_P R_T(p)}{|R_T|}$$

where |*L*| indicates the cardinality (i.e. number of elements) in the set *L*. The ROC curve is produced by varying $t_Q$ between the minimum and maximum values across all data points (i.e. the extrema of set $L\ \cup\ R$).

**Conflict of interests**
The authors have no competing interests to declare.

**Contributions**
GSDG, JJ, TDW and SEB conceived the idea. GSDG and JJ built the endoscope and conducted optical experiments. JJ and TS designed and prepared phase and polarization phantoms. MA and PHJ established the mouse model, prepared tissue specimens and identified early oesophageal lesions. CW designed and fabricated phase and polarization test targets. CRMF took microscope images of mouse samples. AJM implemented aspects of the phase retrieval algorithm. GSDG, JJ, MP, RCF, TDW and SEB analyzed the data. The manuscript was prepared with input from all authors.

**Additional information**
Supplementary information is available with the paper. Correspondence and requests for raw data should be addressed to TDW and SEB.



\begin{titlepage}

\noindent
\Large\textbf{Supplementary Information for:}\\
\\
\Large Quantitative phase and polarisation endoscopy for detection of early oesophageal tumourigenesis\\
\\
\large{George SD Gordon$^{1,2,3}$, James Joseph$^{2,3}$, Maria P Alcolea$^{4}$, Travis Sawyer$^{2,3}$, Alexander J Macfaden$^{1}$, Calum Williams$^{1,2,3}$, Catherine RM Fitzpatrick$^{1,2,3}$, Philip H Jones$^{5}$, Massimiliano di Pietro$^{5}$, Rebecca C Fitzgerald$^{5}$, Timothy D Wilkinson$^{1*}$, Sarah E Bohndiek$^{2,3*}$}\\

\noindent
$^1$ Department of Engineering, University of Cambridge, UK \\
$^2$ Department of Physics, University of Cambridge, UK  \\
$^3$ Cancer Research UK Cambridge Institute, University of Cambridge, UK \\
$^4$ Stem Cell Institute, University of Cambridge, UK \\
$^5$ MRC Cancer Unit, University of Cambridge, UK \\
\\
\large\textbf{Contents}\\
\normalsize
Supplementary Figures \\
Supplementary Note 1: Determination of the fibre bundle transmission matrix\\
Supplementary Note 2: Evaluation of polarisation parameters\\
Supplementary Note 3: Preparation and characterisation of tissue mimicking phantoms\\
Supplementary Note 4: Calculation of entropy and expectation in endoscopic images\\
Supplementary Note 5: Limitations of current design and routes to real-time application\\

\end{titlepage}


\noindent
\Large\textbf{Supplementary Figures}\\

\afterpage{
\thispagestyle{empty}
\begin{figure}[!htp]
	\centering
	\includegraphics[width=0.8\textwidth]{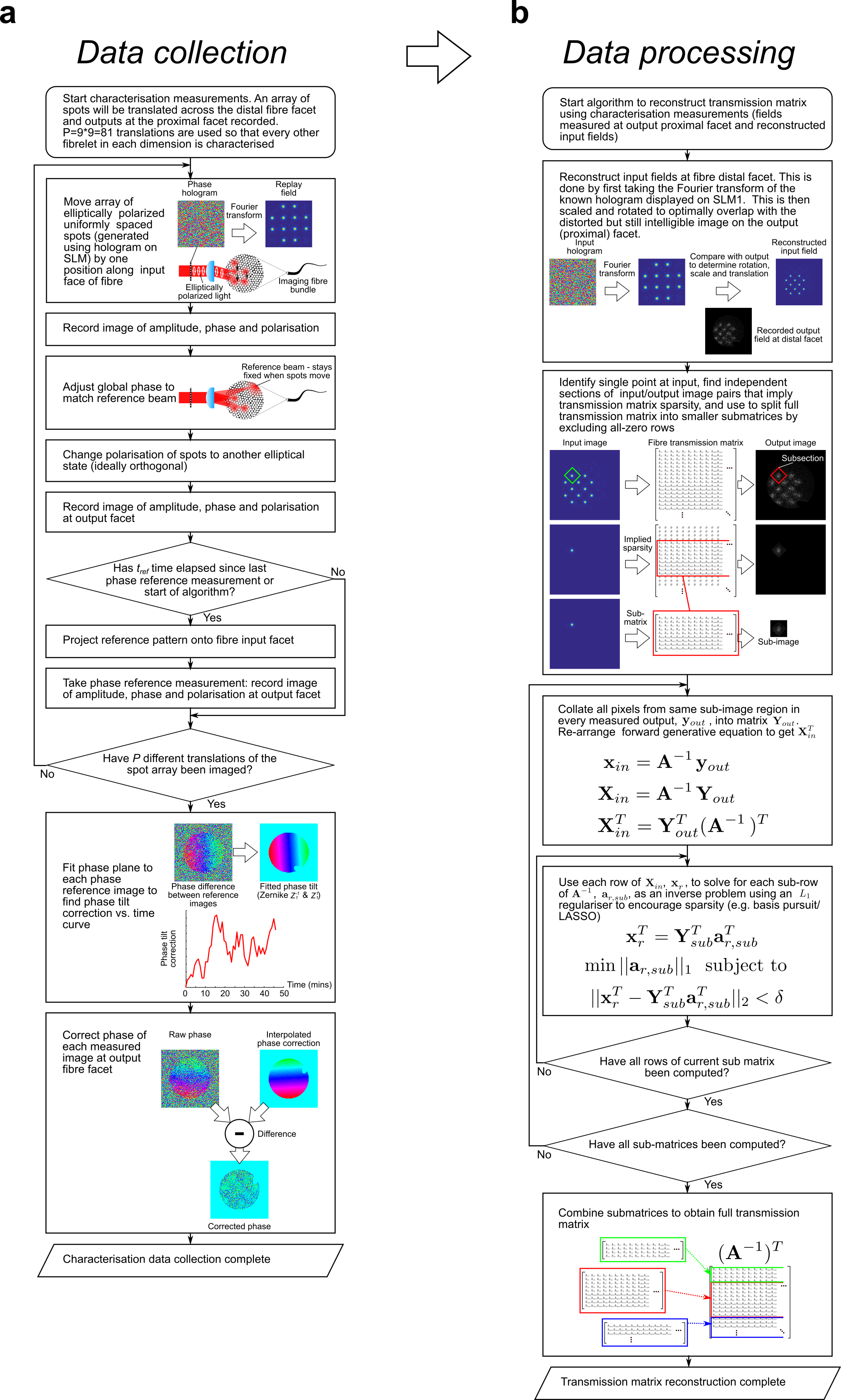}
	\caption{Algorithm detailing how the inverse transmission matrix, $\mathbf{A}^{-1}$ is recovered based on known inputs to the fibre and associated measured outputs: a) Experimental data collection phase. $P = 9 \times 9 = 81$ translations are used so that half the fibrelets in each dimensions can be utilised -- an acceptable trade-off between experimental speed and resolution.  Phase reference measurements are taken every 60 seconds ($t_{ref} = $60 seconds) as experiments show the phase variation on this timescale is small enough to allow interpolation.  b) Data processing performed in software to recover the transmission matrix.}
	\label{fig:matrixRecoveryAlg}
\end{figure}
\clearpage
}

\pagebreak
\begin{figure}[!ht]
	\centering
	\includegraphics[width=1.0\textwidth]{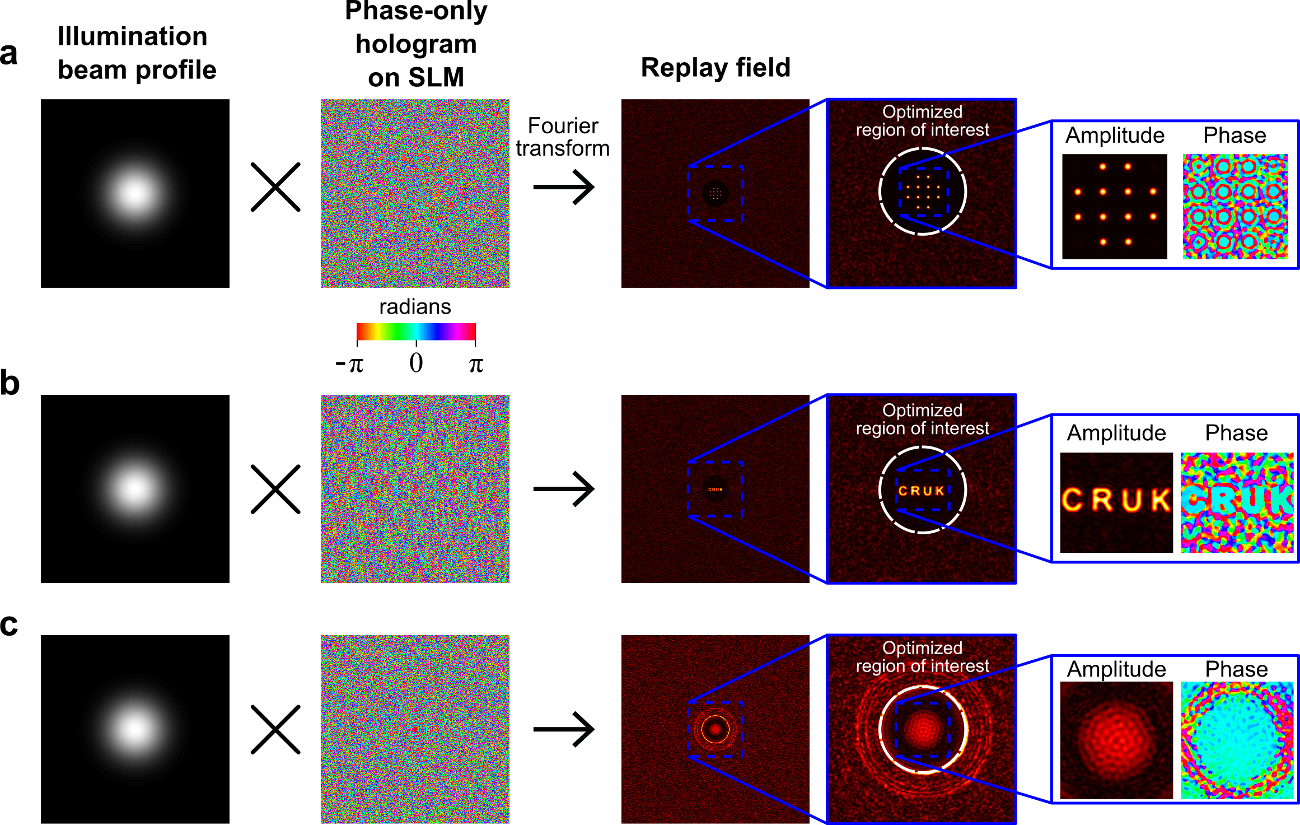}
	\caption{Holograms used for transmission matrix determination, verification, and sample illumination. a) An array of 12 equispaced spots are used to parallelise the fibre characterisation process. Each spot is designed to have an approximately Gaussian profile, however due to the rectangular aperture imposed by the finite extent of the spatial light modulator, in reality they are more like sinc functions with attenuated side lobes. b) Text logo used for verifying reconstruction algorithm. c) Broad, Gaussian amplitude, flat phase illumination profile used for imaging samples. }
	\label{fig:holos}
\end{figure}

\pagebreak
\begin{figure}[!ht]
	\centering
	\includegraphics[width=0.8\textwidth]{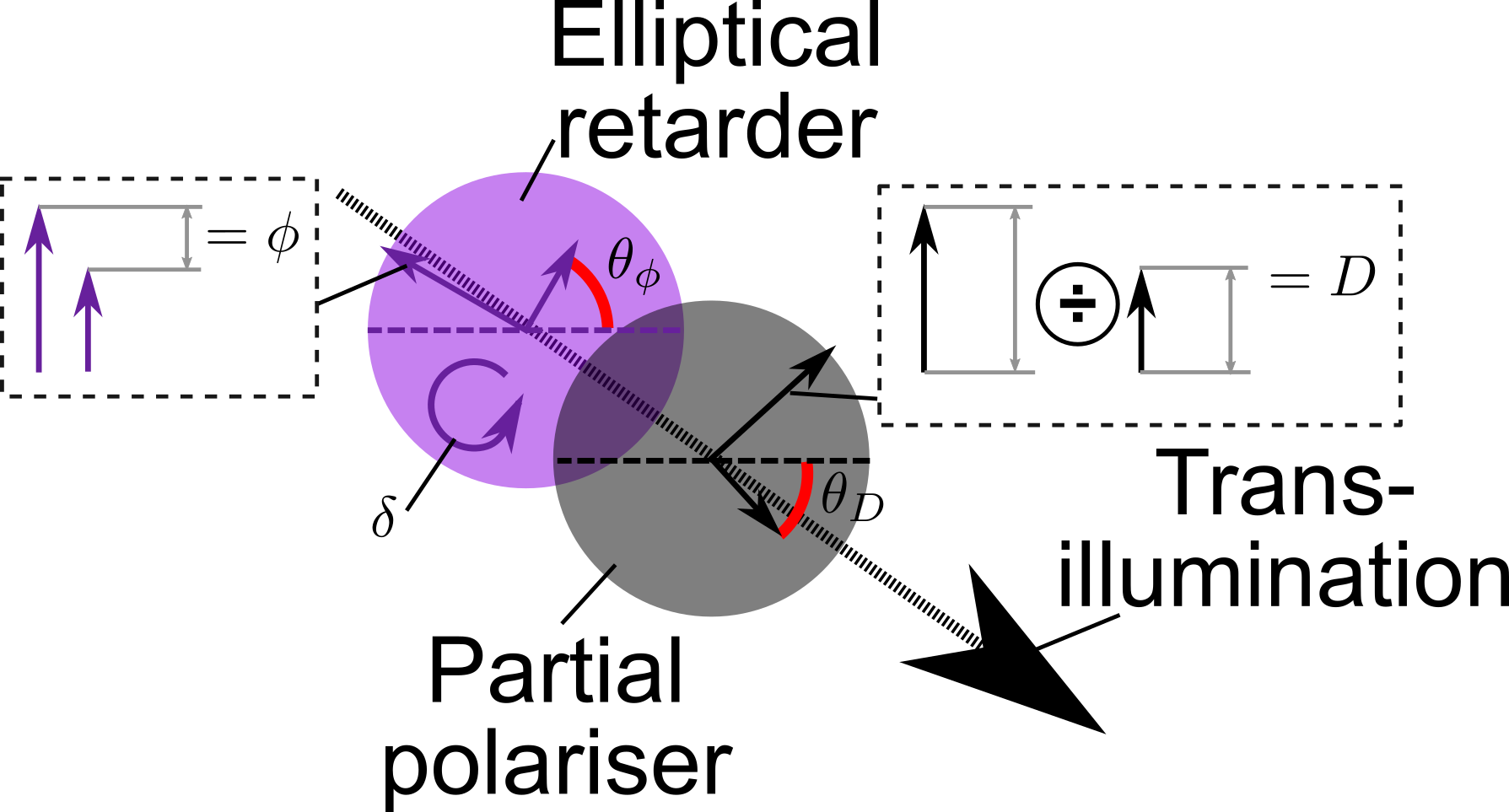}
	\caption{Polarisation model used to factorise Jones matrix at each pixel of sample via Bayesian inference.  The first element is an elliptical retarder that introduces a phase shift $\eta$ (retardance) between the fast and slow optical axes. The retardance axes are oriented at an angle $\theta_{\eta}$ to the horizontal. There is also a component $\xi$ that induces circularity, e.g. representing a chiral molecule. The retarder is followed by a partial linear polariser. The ratio representing the relative power admitted in each of the two axes is $D$ (diattenuation) and the orientation of this diattenuation axis relative to the horizontal is $\theta_D$.}
	\label{fig:polModel}
\end{figure}
\clearpage

\pagebreak
\begin{figure}[!ht]
	\centering
	\includegraphics[width=1.0\textwidth]{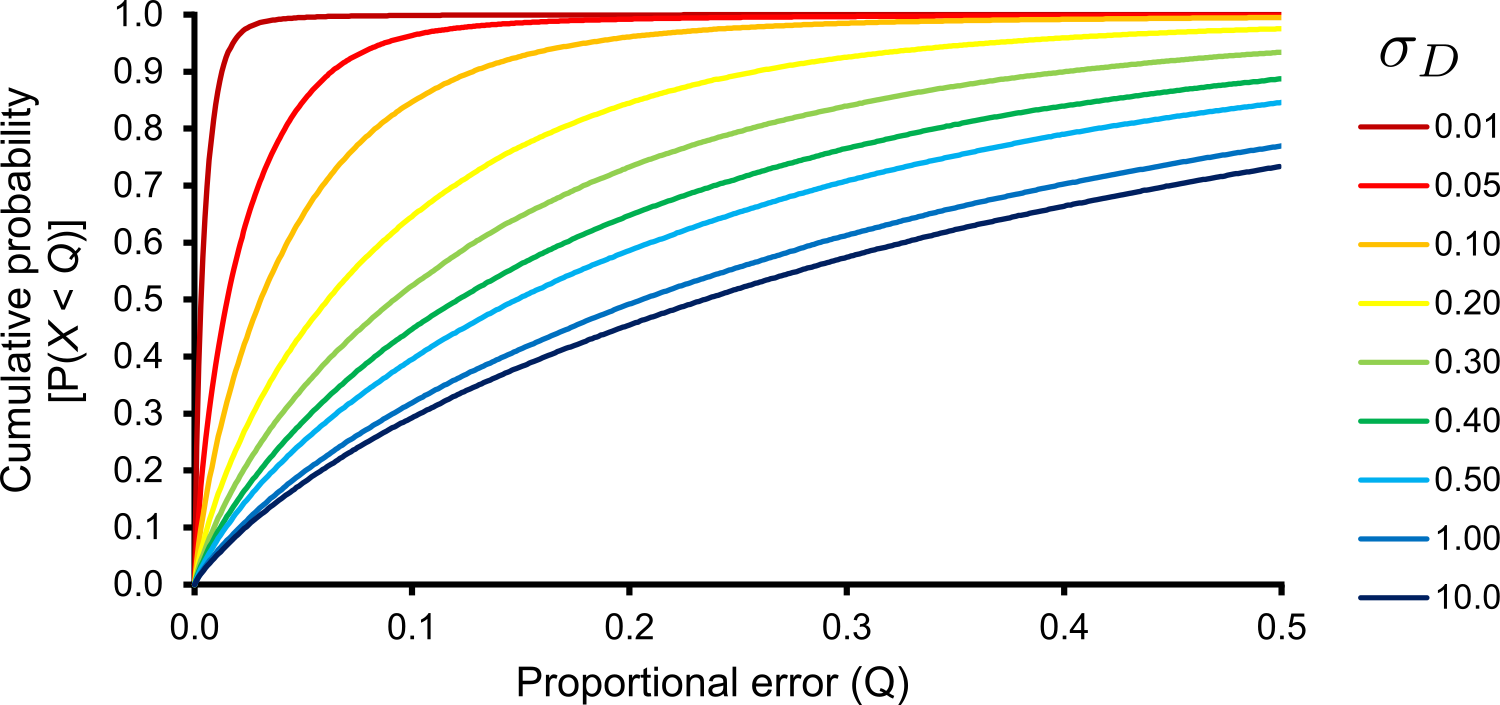}
	\caption{Cumulative distribution function curves showing proportional error incurred in Jones matrix elements when reversing the order of the polariser and retarder of Figure \ref{fig:polModel}.  Curves are generated from Monte-Carlo simulations across 5 polarisation parameters, with diattenuation ($D$) being draw from a normal distribution with different standard deviations ($\sigma_D$).  The measured $\sigma_D$ for the mouse samples used in this work is $<0.01$, which means that in $>99.5$\% of cases the error incurred by assuming a different ordering of polarisation and retarder is $<5$\%. More detail can be found in Supplementary Section \ref{subsec:polOrdering}.}
	\label{fig:polMonteCarlo}
\end{figure}
\clearpage

\pagebreak
\begin{figure}[!ht]
	\centering
	\includegraphics[width=1.0\textwidth]{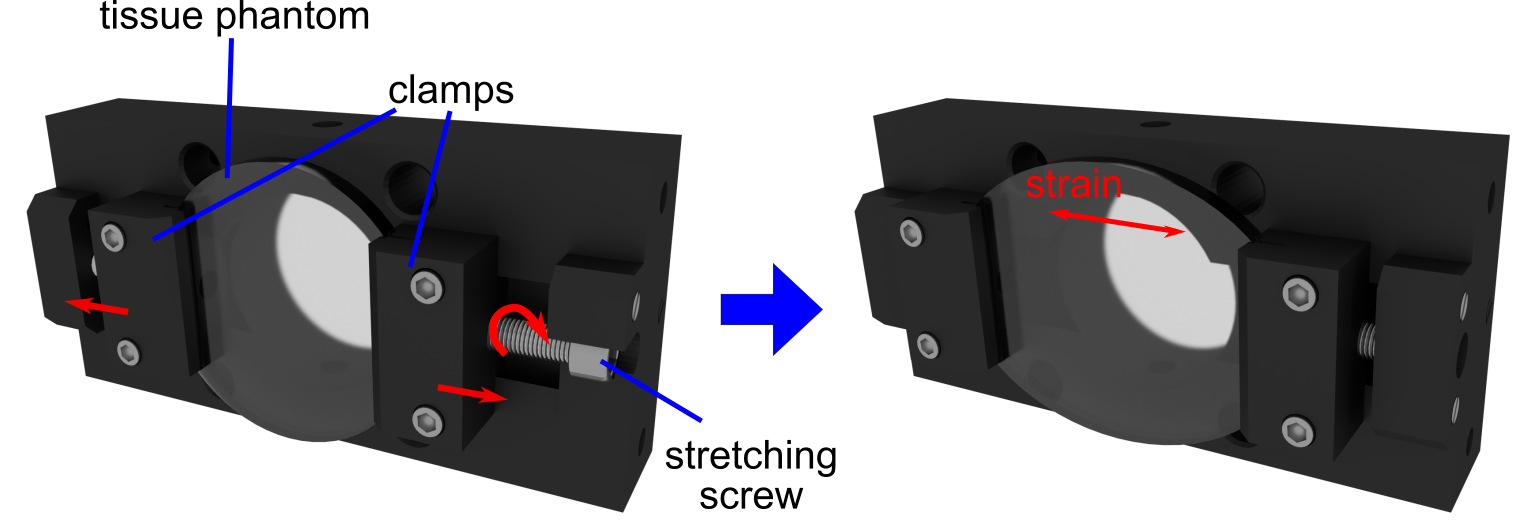}
	\caption{Diagram showing operation of the phantom stretcher assembly used for mounting scattering and birefringent phantoms, and for stretching birefringent phantoms. The phantom is held in place by the clamps, which are securely fastened with screws. The stretching screws (one on either side) are turned in uniform increments to stretch the phantom by increasing amounts in the lateral axis.}
	\label{fig:stretcher}
\end{figure}

\pagebreak
\begin{figure}[!ht]
	\centering
	\includegraphics[width=1.0\textwidth]{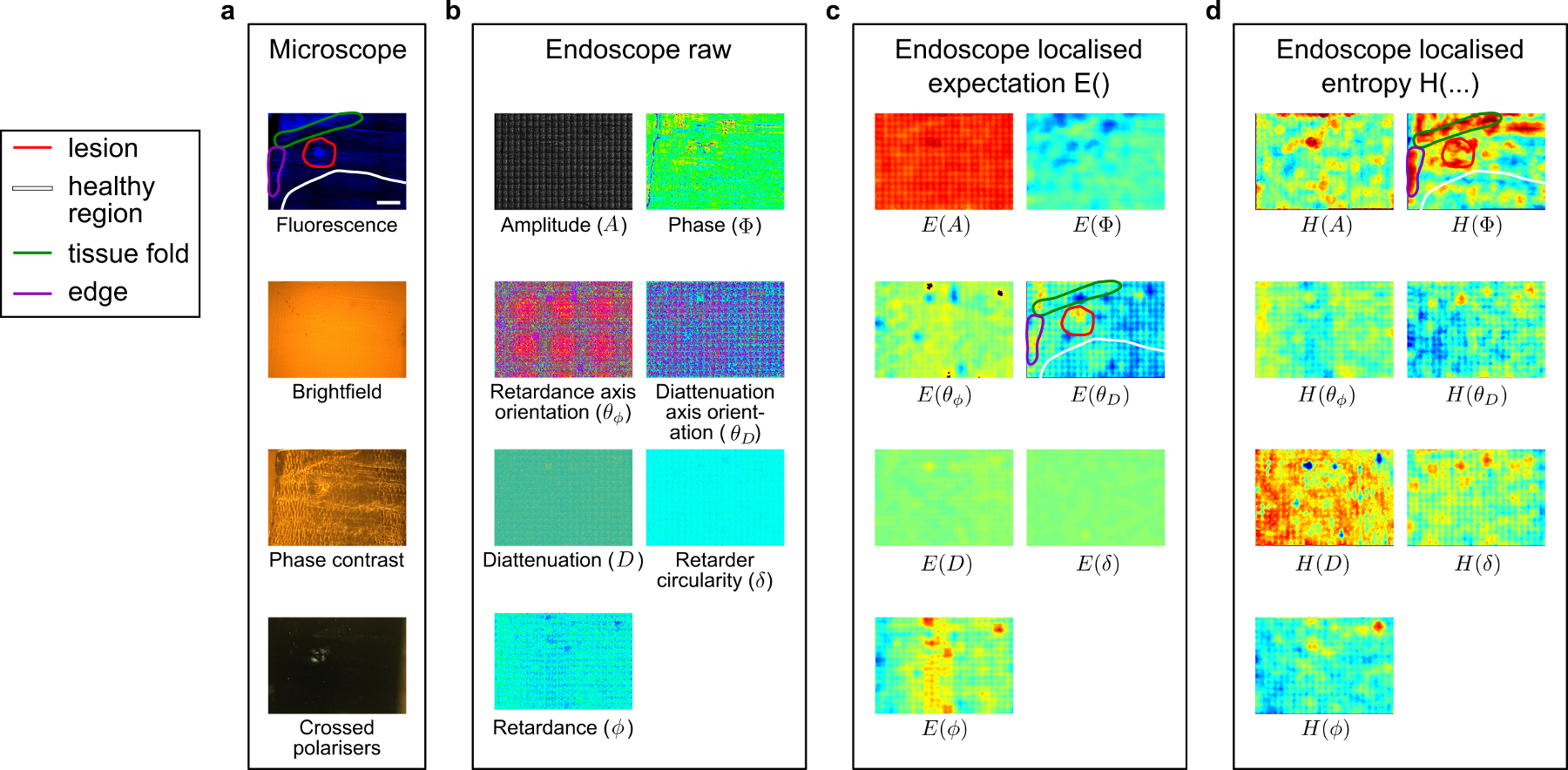}
	\caption{Example data set showing all image modalities for a single mouse sample. Regions of interest are identified based on the legend given on the left. a) Bright field, fluorescence, phase contrast, and crossed-polarisation microscopy of a sample containing a single lesion. The lesion is not visible in the bright field microscope, but is apparent in phase contrast and cross-polarisation microscopy, indicating as expected that phase and polarisation can provide information additional to amplitude-only imaging. b) Raw images from the holographic endoscope, including inferred polarimetic properties. c) Expectation (i.e. spatial average) of the raw quantities. It is noted that in the best performing metric ($\theta_D$) a strong signal is observed at the position of the lesion. d) Entropy of the raw quantities. Phase entropy shows high signal in the lesion area but is also sensitive to scattering caused by a tissue fold (verified in the fluorescence image) and the sample edge. Sample edges are excluded from the analysis presented here, but folds are included as similar features might be expected \emph{in vivo}. Scale bar: 1 mm.}
	\label{fig:fullSample}
\end{figure}

\afterpage{
\thispagestyle{empty}
\begin{figure}[!ht]
	\centering
	\includegraphics[width=0.6\textwidth]{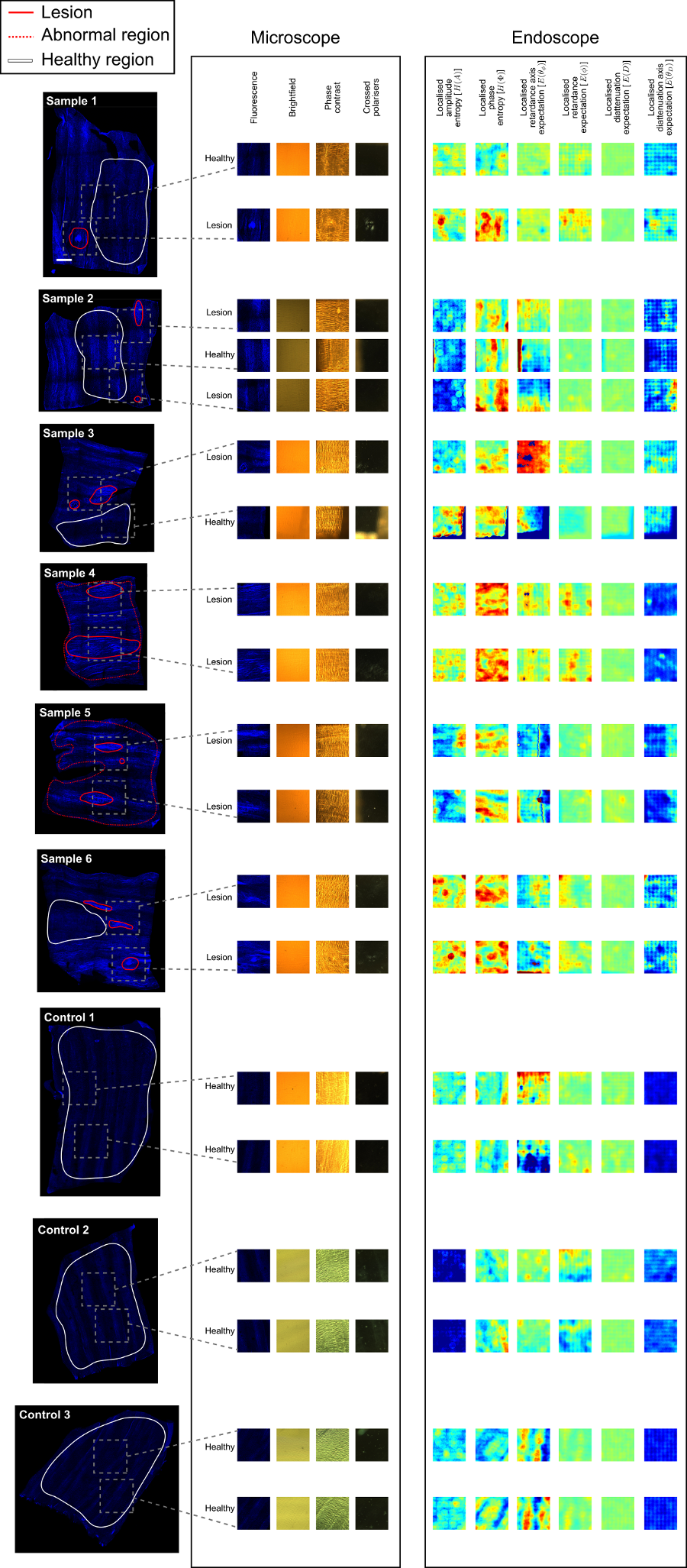}
	\caption{Full field fluorescence images of all 9 samples analysed (6 diseased, 3 control), showing regions designated as healthy, lesion and abnormal by author MPA. The first two categories (healthy and lesion) are used to sample areas for computing the contrast-to-noise ratio. Sections of samples are examined using brightfield, phase contrast and cross-polarised microscopy.  The lesions can be identified much more clearly using phase contrast and crossed-polarised imaging compared to brightfield, in agreement with the hypothesis that phase and polarisation offer additional contrast over amplitude. Processed images of the same regions taken with the endoscope are also shown for comparison.}
	\label{fig:allSamples}
\end{figure}
\clearpage
}

\pagebreak
\begin{figure}[!ht]
	\centering
	\includegraphics[width=0.55\textwidth]{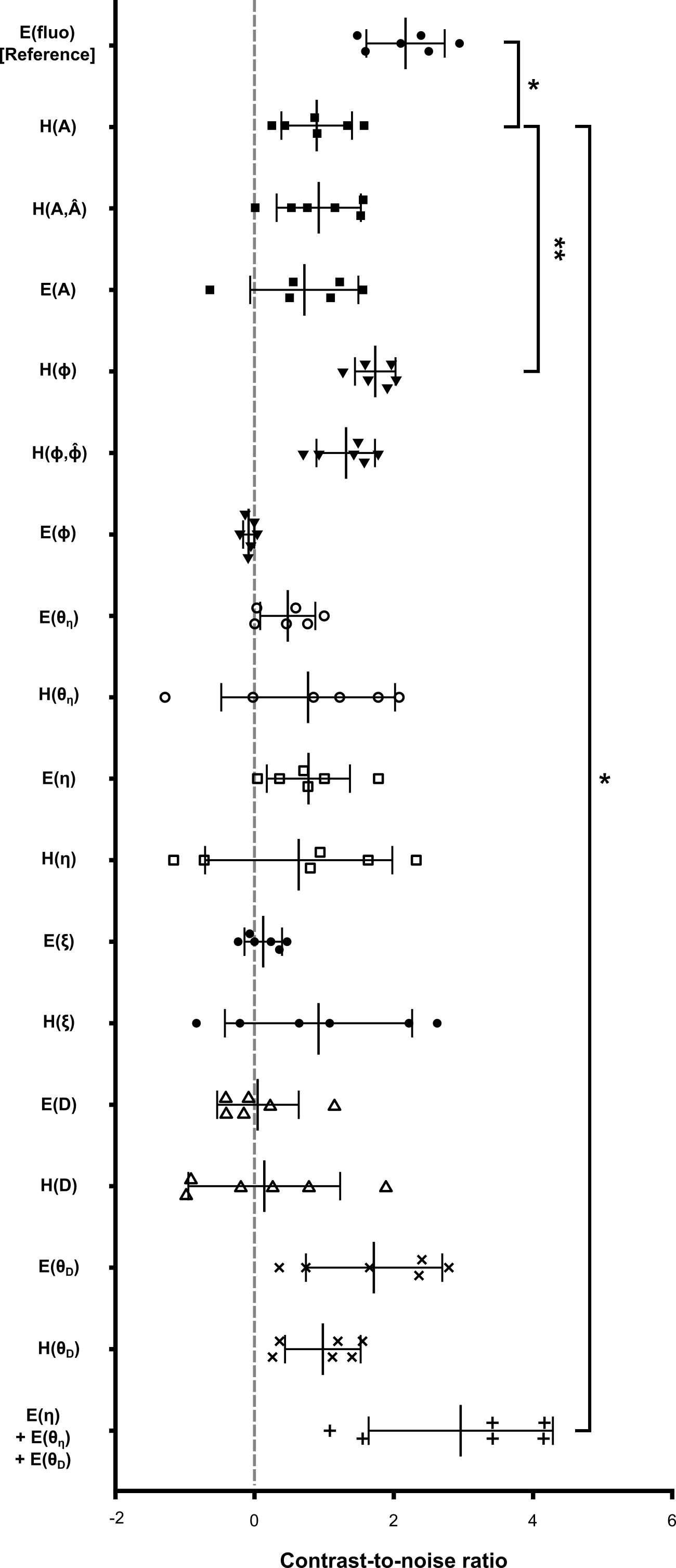}
	\caption{Contrast-to-noise ratio for all phase and resolved polarimetric properties using both spatial entropy, $H()$, and expectation, $E()$. $H(A,\hat{A})$ and $H(\phi,\hat{\phi})$ represent co-occurrence amplitude and phase entropy respective (described in Supplementary Figure \ref{fig:entropyCalcAlg} but are not significantly different to the standard entropy metric and are thus not used for further analysis.}
	\label{fig:CNRall}
\end{figure}
\clearpage

\pagebreak
\begin{figure}[!ht]
	\centering
	\includegraphics[width=0.5\textwidth]{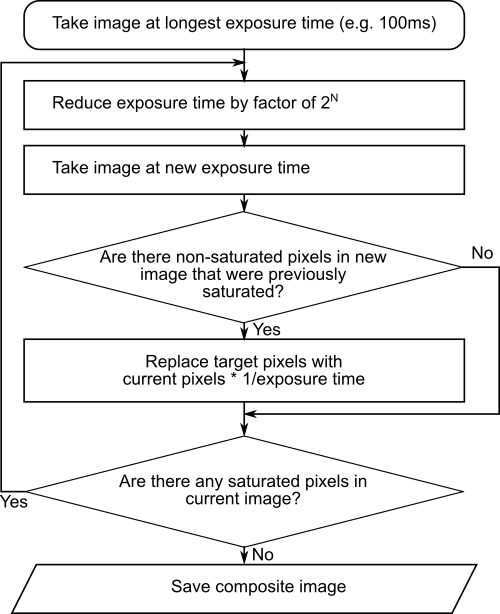}
	\caption{Algorithm for measuring high dynamic range images using adjustable exposure time. The exposure time is reduced at each step by a factor of $2^N$, where $N \in \{1\cdots8\}$ for an 8-bit camera. Here we set $N=1$ so each pixel displays the same percentage error, since every saturated pixel at the longest exposure will ultimately be recorded with 8-bit precision for the 8 most significant bits. }
	\label{fig:HDR}
\end{figure}

\afterpage{
\thispagestyle{empty}
\begin{figure}[!htp]
	\centering
	\includegraphics[width=0.8\textwidth]{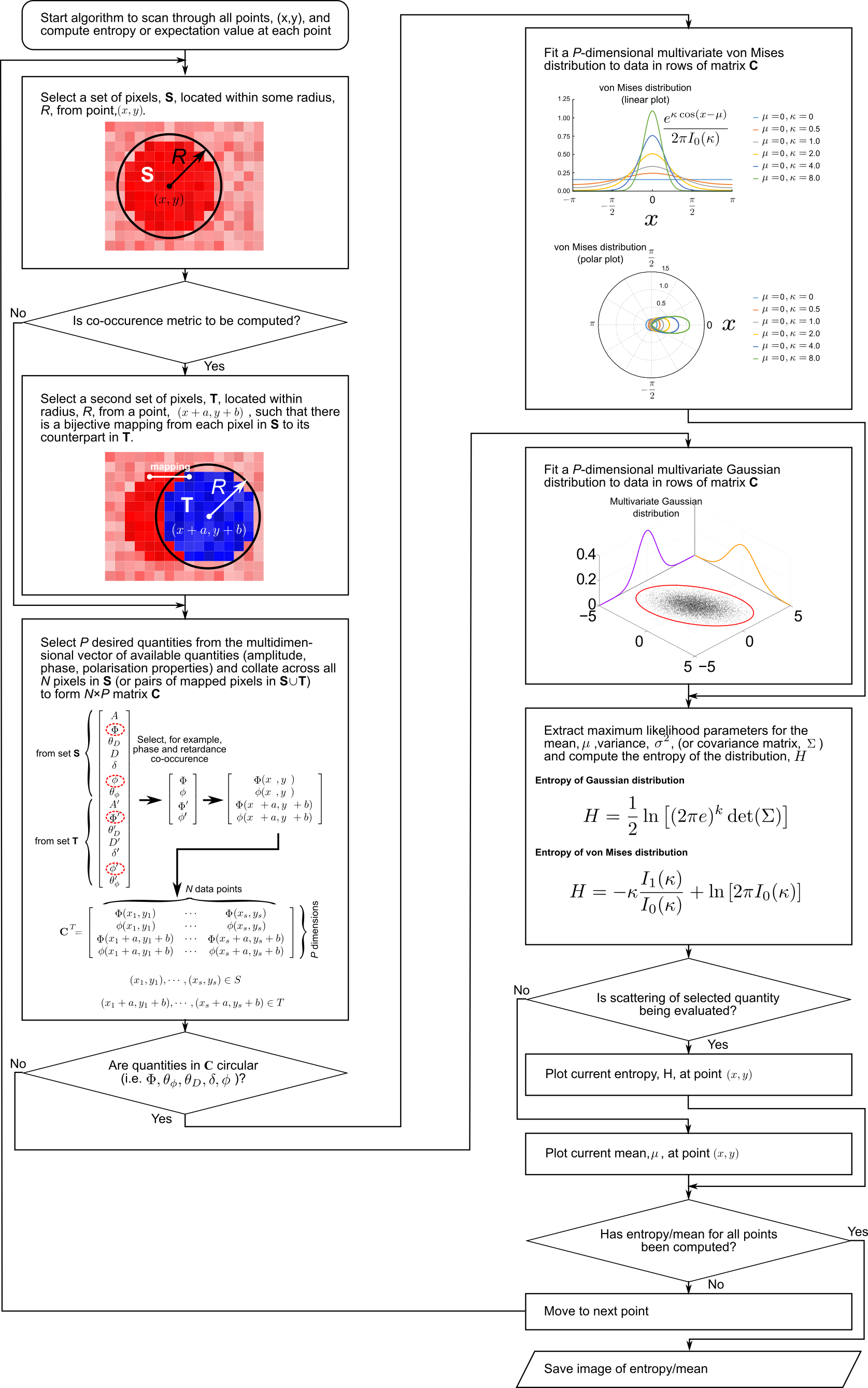}
	\caption{Algorithm detailing how entropy and expectation metrics are computed from raw images of amplitude, phase and polarisation.  The value of $R$, the size of the filter window, is chosen to be 15 because it is sufficiently narrow so as not to `smooth out' the smallest lesions ($\sim$30 pixels in size in this dataset), but offers sufficient data points to reliably fit a 2D distribution (e.g. for co-occurrence analysis). The co-occurence metric computed here uses an offset vector of $(a,b)=(0,4)$.  This is small enough so most pixels that are compared belong to the same type of tissue (e.g. lesion) but large enough that decorrelation might be expected. These parameters were also swept and it was observed that small changes in them did not significantly affect results, nor did the direction of the co-occurrence offset vector.}
	\label{fig:entropyCalcAlg}
\end{figure}
\clearpage
}

\pagebreak
\begin{figure}[!ht]
	\centering
	\includegraphics[width=1.0\textwidth]{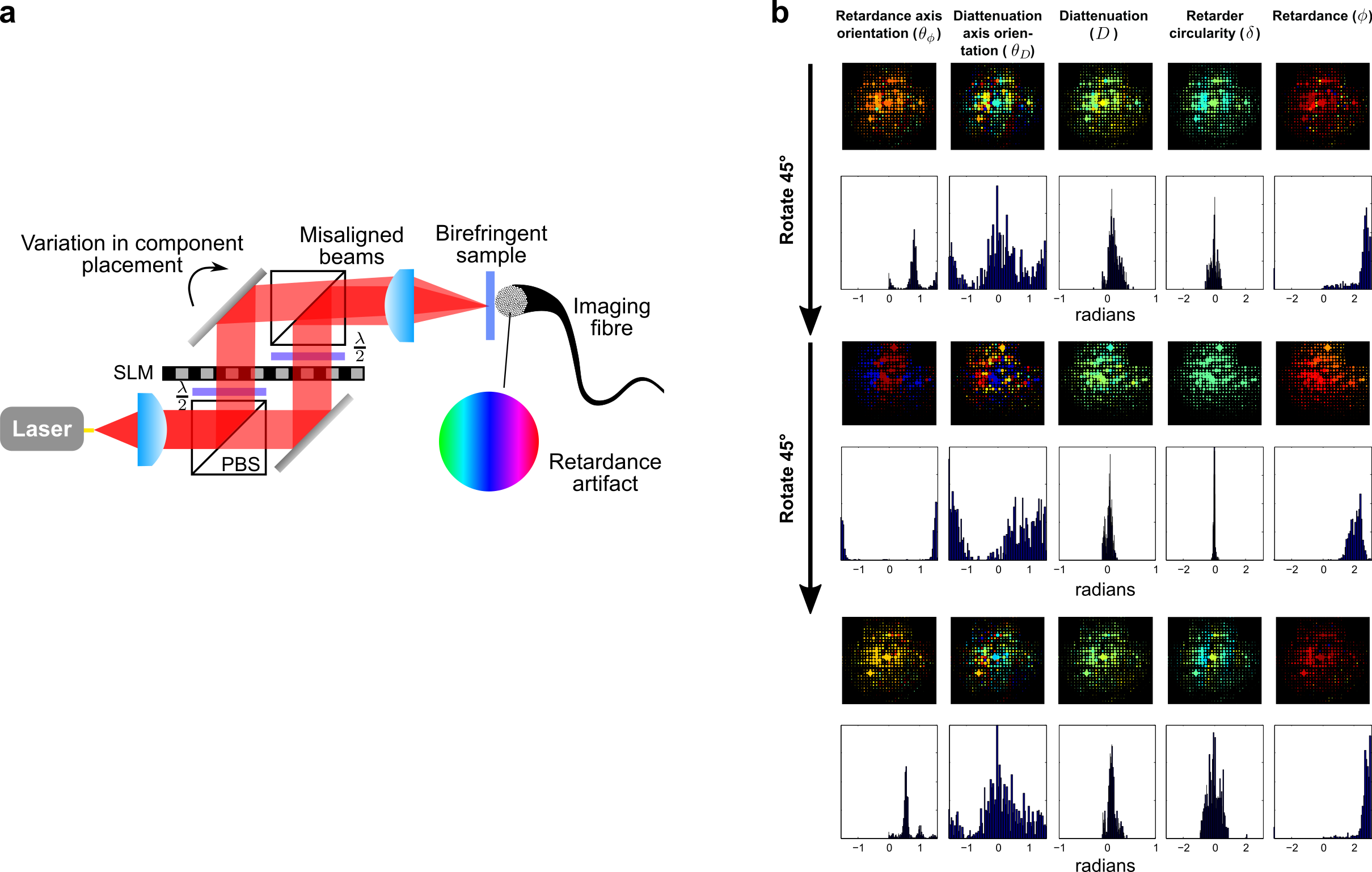}
	\caption{Correction of polarisation misalignment artefacts: a) Small variations between polarisation arms can result in the beams not being parallel.  This is not noticeable if the sample retardance axis orientation ($\theta_{\eta}$) is aligned with the polarisation axis of the two arms.  However, for a general birefringent sample with an arbitrary $\theta_{\eta}$ the misaligned beams are cross-coupled creating a phase-tilt artefact in the recovered retardance. b) This artefact can be effectively compensated by re-expressing the problem in a linear polarisation basis at angle $\theta_\eta$ to the illumination polarisation axis and performing joint inference on neighbouring pixels. This is experimentally demonstrated using a rotating half-waveplate and inferring the 5 polarisation parameters for our model. As expected, the retardance ($\eta$) stays constant but the angle of the retardance axis ($\theta_{\eta}$) changes when the waveplate is rotated. Furthermore, no spatially varying retardance is observed in the images demonstrating successful removal of artefacts.}
	\label{fig:waveplateImaging}
\end{figure}
\clearpage

\pagebreak
\noindent
\Large\textbf{Supplementary Note 1:}\\
\Large\textbf{Determination of the fibre bundle transmission matrix}\\
\normalsize
\setcounter{section}{1}

\subsection{Mathematical framework of the forward problem}
\label{subsec:tMatRecovery}
If we define the electric field of light incident on the distal facet of the fibre bundle at a point $(u,v)$ as $T_i (u,v)$ and the electric field incident on the camera sensor at a point $(x,y)$ as $T_o (x,y)$, then we can model the relationship between the two using a spatially dependent point-spread function (or \emph{Green's function}), $h(u,v,x,y)$ as:

\begin{equation}
\label{eq:greenFunc}
T_o(x,y) = \iint h(u, v, x, y) T_i(u,v) \mathrm{d}u \mathrm{d}v
\end{equation}

In the case of free-space propagation, $h(u,v,x,y)$ can be set as $\frac{e^{ikr}}{i r\lambda}$ to give the Fresnel diffraction integral.  However, in our case $h(u,v,x,y)$ is a more complex function that is not known \emph{a priori}. A simple approach to approximate this is to discretise the inputs and outputs and model $h(u,v,x,y)$ with a matrix. To achieve this, we let $T_i(u,v)$ be discretised and then lexicographically ordered into an $n \times 1$ column vector, $\mathbf{x}_h \in \mathbb{C}^n$. We then let $T_o(x,y)$ be discretised and lexicographically ordered into an $m \times 1$ column vector, $\mathbf{y}_h \in \mathbb{C}^m$. In this latter case, an obvious choice of sampling scheme would be to use the values recorded by each pixel of the image sensor. Discretising $h(u,v,x,y)$ leads to the matrix equation:

\begin{equation}
\mathbf{y}_h = \mathbf{A}_{hh} \mathbf{x}_h
\end{equation}

\noindent where $\mathbf{A}_{hh}$ is an $m \times n$ transfer matrix mapping horizontally polarised light at the input to horizontally polarised light at the output.  Repeating the discretisation for the vertically polarised input and output fields, then combining we find:

\begin{equation}
\mathbf{y} =  \left[ \begin{array}{c}
								\mathbf{y}_h \\
								\mathbf{y}_v \end{array} \right]
\end{equation}

\begin{equation}
\mathbf{x} =  \left[ \begin{array}{c}
								\mathbf{x}_h \\
								\mathbf{x}_v \end{array} \right]
\end{equation}

\begin{equation}
\mathbf{A} =  \left[ \begin{array}{cc}
								\mathbf{A}_{hh} & \mathbf{A}_{hv}\\
								\mathbf{A}_{vh} & \mathbf{A}_{vv} \end{array} \right]
\end{equation}

\begin{equation}
\label{eq:forwardEq}
\mathbf{y} = \mathbf{A} \mathbf{x}
\end{equation}

\noindent where $\mathbf{A}$ is the full $2m \times 2n$ transfer matrix.  In practice we record images of the proximal facet of the MCF in the plane of the camera ($\mathbf{y}$) and use these to reconstruct the input fields in the distal facet plane ($\mathbf{x}$).  Equation \ref{eq:forwardEq} can therefore be considered a linear inverse problem: for a known $\mathbf{y}$ and $\mathbf{A}$, the goal is to find $\mathbf{x}$.  One method of solving this problem is to find a pseudo-inverse of $\mathbf{A}$ such that:

\begin{equation}
\mathbf{x} = \mathbf{A}^{-1} \mathbf{y}
\end{equation}

However, we do not typically know $\mathbf{A}^{-1}$ in advance and so we must first determine this before $\mathbf{y}$ can be recovered.  Consider now combining $p$ vectors representing different input fields at the distal facet to form a matrix, $\mathbf{X}$:

\begin{equation}
\label{eq:xMat}
\mathbf{X} =  \left[ \begin{array}{cccc}
\mathbf{x}_1 & \mathbf{x}_2 & \cdots &\mathbf{x}_p
 \end{array} \right]
\end{equation}

For each of these $p$ vectors, the field exiting the distal facet is recorded at the camera plane, creating another matrix, $\mathbf{Y}$:

\begin{equation}
\label{eq:yMat}
\mathbf{Y} =  \left[ \begin{array}{cccc}
\mathbf{y}_1 & \mathbf{y}_2 & \cdots &\mathbf{y}_p
 \end{array} \right]
\end{equation}

Collectively, $\mathbf{X}$ and $\mathbf{Y}$ are termed the \emph{calibration measurements}.  We can then write the following:

\begin{equation}
\mathbf{X} = \mathbf{A}^{-1} \mathbf{Y}
\end{equation}

\begin{equation}
\label{eq:recoveryEq}
\mathbf{X}^T = \mathbf{Y}^T (\mathbf{A}^{-1})^T
\end{equation}

By solving for each of the $p$ columns of $(\mathbf{A}^{-1})^T$ using the respective column of $\mathbf{X}^T$, we can reduce the problem of finding the inverse transmission matrix to $p$ linear inverse problems. Solving each of these provides a single row of $\mathbf{A}^{-1}$:

\begin{equation}
\label{eq:recoveryEq_sing}
\mathbf{x}_r^T = \mathbf{Y}^T \mathbf{a}_r^T
\end{equation}

\noindent where $\mathbf{x}_r$ is the $r^{th}$ row of $\mathbf{X}$ and $\mathbf{a}_r$ is the $r^{th}$ row of $\mathbf{A}^{-1}$.

\subsection{Experimental measurements}
\label{subsubsec:parallelisation}
The full process of producing appropriate sets of inputs and outputs, $\mathbf{X}$ and $\mathbf{Y}$, for recovering $\mathbf{A}^{-1}$ is detailed in Supplementary Figure \ref{fig:matrixRecoveryAlg}. We control the vectors $\mathbf{x}_i$ of Equation \ref{eq:xMat} for $i=1\cdots p$ by displaying an appropriate hologram on SLM1. In this case, each $\mathbf{x}_i$ represents one of 81 translations (over a $9\times9$ grid) in 9$\mu$m steps (limiting spatial resolution) of an array of 12 equispaced spots (Supplementary Figure \ref{fig:holos}a) in one of 3 elliptical polarisation states, resulting in a total of $p=81 \times 3 = 243$ inputs.  Elliptical states are used so as to maximise power transmitted through the fibre, thereby minimising the effect of noise compared to linear states with only half the power.  A minimum of two states are required to resolve the 2 orthogonal polarisation components but further states can reduce noise.  Using 3 states produced a $\sim 33\%$ reduction in noise, but increasing to 4 states provided little improvement as the noise performance was observed to be limited by approximation noise in the basis pursuit algorithm and non-flatness of the hologram.  3 states are therefore selected as a balance between noise reduction and experimental time reduction.  The translation step size is chosen so that every second fibrelet in each dimension is utilised.  This offers a balance between reducing experimental time ($<$1 hour) while still offering sufficient resolution for examining small lesions (size $>50\mu$m).

The choice of 12 spots ensures individual spots are spaced sufficiently far from each other that there is negligible interference between them at the proximal facet of the fibre bundle (even after been scrambled by the matrix $\mathbf{A}$) and they can be considered to behave independently. In a pixel basis representation, the resulting fibre bundle transmission matrix is both \emph{sparse}, and in some sense \emph{local} (technically, a \emph{band matrix}). Hence, not only are most entries zero, but the non-zero entries are clustered together. Consequently, a single image containing 12 spots can be split up to produce 12 separate characterisation images. Each image simply uses the relevant subsection of the full image and sets all other pixels to zero. This means that there are effectively $12 \times 243 = 2916$ characterisation images from only 243 actual measurements. Each of the 81 spatial translations effectively provides 12 pixels of resolution at the distal facet (one per spot).  Therefore, in total there are $12 \times 81 = 972$ pixels that can be spatially resolved at the distal facet.

When using a phase retrieval algorthm (see Online Methods), the phase recovered for each image has an arbitrary global offset. If measured na\"{\i}vely, the columns of $\mathbf{Y}$ in Equation \ref{eq:yMat} would have an arbitrary phase offset between them. To compare different phase images, a fixed phase reference is created during the characterisation measurements by displaying a superposition of two holograms on SLM1: the first hologram creates the translatable array of spots and the second creates a single spot that is kept in a fixed position. The fixed position is a set of fibrelets at the edge of the MCF, as shown in Supplementary Figure \ref{fig:matrixRecoveryAlg}a. Recovered phase values are normalised to the phase of this spot at the proximal facet of the fibre. This is effectively a form of common-path interferometry. Such a global phase offset is also termed \emph{piston} and can be thought of as a phase surface polynomial of order 0 \cite{Loterie2015a}. Here, we observed experimentally that in addition to this 0$^{th}$ order correction, a 1$^{st}$ order time-dependent phase correction,termed \emph{phase tilt}, was also required. The origin of the time-dependent phase tilt is likely due to temperature, affecting the SLM and optical mounts \cite{Garcia-Marquez2011}. Air currents could also play a role, however the experiment is entirely contained in an optical box so their effect should be negligible. The 1$^{st}$ order phase correction term is tracked by measuring a reference image at 60 second intervals during characterisation. The relative phase tilt between reference images is computed and interpolation with respect to time is then used to estimate the correction to be applied to characterisation measurements, $\mathbf{y_i}$.

\subsection{Solving the inverse problem}
The region of the image sensor used for data collection is 1200$\times$1200 pixels and so the inverse transmission matrix, $\mathbf{A}^{-1}$, has 2$\times$1200$\times$1200 =  2.88 million columns when the two polarisation states are considered. Similarly, the 81 translations of 12 spots in three polarisation states at the proximal facet gives a total of 2916 rows for $\mathbf{A}^{-1}$. To store this 2916 $\times$2880000 matrix at single floating point precision would require $>$30Gb of memory. This can make recovery of the matrix and subsequent operations infeasibly slow.

However, because the transmission matrix is sparse and the few non-zero elements are grouped together (i.e. a banded structure as discussed above), it is possible to reconstruct only the parts of $\mathbf{A}^{-1}$ where the non-zero elements are located. Consider again Equation \ref{eq:recoveryEq_sing} in which we seek to solve for row $r$ of $\mathbf{A}^{-1}$, $\mathbf{a}_r$: 

\begin{equation}
\mathbf{x}_r^T = \mathbf{Y}^T \mathbf{a}_r^T
\end{equation}

Here, $\mathbf{x}_r$ is a single row of $\mathbf{X}$ and thus corresponds to a single spatial position across all input calibration fields at the distal facet.  Similarly, $\mathbf{a}_r$ is a single row of $\mathbf{A}^{-1}$ and therefore represents the coupling to the single spatial position represented by $\mathbf{x}_r$ from all spatial positions at the proximal facet, represented columns of $\mathbf{Y}$.  However, because we know $\mathbf{A}^{-1}$ is sparse and banded, the only contributions to $\mathbf{x}_r$ will come from points in a confined region on the distal facet.  The distal and proximal facet coordinate axes are the same as a result of an affine transform (see Online Methods) and so if the coordinates of the spatial position of $\mathbf{x}_r$ are $(a,b)$, then the relevant points on the distal facet will be neighbouring $(a,b)$.  We therefore select a set of spatial points on the distal facet (i.e. columns $\mathbf{Y}$) that are located within a square centred on $(a,b)$ with side length 68.2$\mu$m.  This length is exactly equal to the spacing between adjacent spots in the array used to characterise the fibre (Supplementary Figure \ref{fig:holos}a), which was in turn set so as to avoid interference of spots at the distal facet (Online Methods). This process is illustrated in Supplementary Figure \ref{fig:matrixRecoveryAlg}.  Equation \ref{eq:recoveryEq_sing} then becomes:

\begin{equation}
\label{eq:subEq}
\mathbf{x}_{r}^T = \mathbf{Y}_{sub}^T \mathbf{a}_{r,sub}^T
\end{equation}

\noindent where $\mathbf{Y}_{sub}$ is the selected subset of rows of $\mathbf{Y}$ and 
 $\mathbf{a}_{r,sub}$ is the corresponding subset of columns of $\mathbf{A}^{-1}$. Typically, $\mathbf{a}_{r,sub}$ has around 60,000 columns over 2 polarisations, a 50-fold reduction from the original 2.88 million.  However, given that $\mathbf{x}_r$ has only $\sim$3000 columns, the problem is underdetermined by a factor of 10.  The problem can still be solved by exploiting the prior knowledge that each $\mathbf{a}_{r,sub}$ ought to be sparse because spatial positions towards the edge would likely couple little or no power directly from the central positions (in agreement with observations). This is achieved using an $L1$ regulariser so that the solution to Equation \ref{eq:subEq} becomes \cite{Zhang2015}:
 
\begin{equation}
\min ||\mathbf{a}_{r,sub}||_1 \mathrm{~~subject~to~~} ||\mathbf{x}_r^T - \mathbf{Y}_{sub}^T \mathbf{a}_{r,sub}^T||_2 < \delta
\label{eq:solver}
\end{equation}

\noindent where $\delta$ is a parameter that can be adjusted to favour a more or less sparse solution.  Equation \ref{eq:solver} is solved numerically using a basis pursuit denoising algorithm implemented in the SPGL1 package \cite{spgl1:2007}.

\pagebreak
\noindent
\Large\textbf{Supplementary Note 2:}\\
\Large\textbf{Evaluation of polarisation parameters}\\
\normalsize
\setcounter{section}{2}
\setcounter{subsection}{0}
\subsection{Polarisation model}
\label{subsec:polModel}
When polarised light interacts with the turbid medium of tissue, it may simultaneously encounter: depolarisation, arising from multiple scattering of sub-cellular structures; birefringence and diattenuation, arising from anisotropic muscle fibres and structural proteins; and circularity caused by chiral molecules and structures. 

The propagation of polarised light and its interaction within a sample can be described by either Jones (field-based) or Stokes-Mueller (intensity-based) calculus \cite{Simon1982}. Stokes-Mueller is more commonly used for tissue polarimetry as it can handle partial polarisation states, represented by the depolarisation parameter  \cite{Ghosh2011,Qi2016,Yao1999,Pierangelo2013a}, though Jones calculus has found use in polarisation-sensitive optical coherence tomography of tissue \cite{Park2009,Li2015}.  However, if the coherence length of the light source is substantially longer than possible scattering paths in the tissue, then the light always has a well-defined polarisation state \cite{Simon1982}.  This corresponds to a light source with a high degree of temporal coherence, i.e. monochromatic light. The laser used here has a coherence length of around 10 m compared to the typical transport mean free path for tissue of $<$ 1 mm \cite{Ntziachristos2010} and so this condition is satisfied. In addition to this temporal coherence, the light field must also exhibit spatial coherence in order to have a well-defined polarisation state.  This can be violated, for example, if the integration area of each camera pixel contains  different polarisation states.  However, in our case the light field from the sample is first passed through the fibre bundle where the single-mode propagation in each fibrelet admits only a single polarisation state -- a filter of sorts.  Each fibrelet is imaged on the camera using $>$10 pixels so it can therefore be assumed that the polarisation state is uniform within each pixel area. Depolarisation is therefore considered to be negligible in this work.

In order to examine the properties of birefringence, diattenuation (also called \emph{linear dichroism}) and circularity, we choose a generalised decomposition of the Jones matrix into an arbitrary elliptical retarder followed by a partial linear polariser. The elliptical retarder accounts for birefringent retardance and circularity. The partial linear polariser accounts for the fact that anisotropic molecules might absorb light preferentially along some axis. This is illustrated in Supplementary Figure \ref{fig:polModel}. It is noted that this model does not account for some polarimetric properties, notably circular dichroism. Circular dichroism has applications in spectroscopic analysis of chiral molecules (e.g. many proteins and biological molecules) but produces weak signals so is typically only effective for molecules in concentrated solution \cite{Fasman1996}. Further, little or no work has been produced to date linking circular dichroism at optical wavelengths to imaging of cancer. Therefore, for simplicity it is neglected in this work.

Having measured a set of $n$ output polarisation vectors from the sample $\mathbf{V}(x,y) = \left[\mathbf{v}_1(x,y) \cdots \mathbf{v}_n(x,y)\right]$ and knowing the corresponding $n$ sample illumination polarisation vectors $\mathbf{U}(x,y) = \left[\mathbf{u}_1(x,y) \cdots \mathbf{u}_n(x,y)\right]$ the 2$\times$2 complex Jones matrices $\mathbf{J}(x,y)$ representing how the sample maps input polarisation state to output polarisation state at each pixel can be determined through:

\begin{equation}
\label{eq:forwardModelJones}
\mathbf{V}(x,y) = \mathbf{J}(x,y) \mathbf{U}(x,y)
\end{equation}

\noindent where $\mathbf{J}(x,y)$ is the Jones matrix of a particular pixel at $(x,y)$. We express our decomposition of $\mathbf{J}(x,y)$ as:

\begin{equation}
\label{eq:jonesMat}
\mathbf{J}(x,y) = \mathbf{A}_{pol} \mathbf{A}_{ret}
\end{equation}

\begin{equation}
\label{eq:polMat}
\mathbf{A}_{pol} = \left( \begin{array}{cc}
\cos \theta_D & \sin \theta_D \\
-\sin \theta_D & \cos \theta_D \end{array} \right)
\left( \begin{array}{cc}
\sqrt{1+D} & 0 \\
0 & \sqrt{1-D} \end{array} \right)
\left( \begin{array}{cc}
\cos \theta_D & -\sin \theta_D \\
\sin \theta_D & \cos \theta_D \end{array} \right)
\end{equation}

\begin{equation}
\label{eq:retMat}
\mathbf{A}_{ret} = \left( \begin{array}{cc}
1 & 0 \\
0 & e^{i\xi}\end{array} \right)
\left( \begin{array}{cc}
\cos \theta_{\eta} & \sin \theta_{\eta} \\
-\sin \theta_{\eta} & \cos \theta_{\eta} \end{array} \right)
\left( \begin{array}{cc}
e^{i\eta/2} & 0 \\
0 & e^{-i\eta/2} \end{array} \right) \cdot \left( \begin{array}{cc}
\cos \theta_{\eta} & -\sin \theta_{\eta} \\ \sin \theta_{\eta} & \cos \theta_{\eta} \end{array} \right)
\left( \begin{array}{cc}
1 & 0 \\
0 & e^{-i\xi}\end{array} \right)
\end{equation}

\noindent allowing us to recover the following 5 polarisation properties: diattenuation, $D$, diattenuation axis orientation, $\theta_D$, retarder circularity, $\xi$, retardance, $\eta$, and retardance axis orientation, $\theta_{\eta}$.  From these equations, it is observed that the inferred parameters are all real numbers limited to the following ranges:

\begin{equation}
\label{eq:paramRanges}
\theta_{\eta} \in (-\pi/2, \pi/2],~~~~~~~\eta \in (-\pi, \pi],~~~~~~~\xi \in (-\pi, \pi], ~~~~~~~D \in [-1, 1],~~~~~~~\theta_D \in (-\pi/2, \pi/2]
\end{equation}

\subsection{Ordering of polarisation elements}
\label{subsec:polOrdering}
The model for inference of polarimetric parameters, introduced in Section \ref{subsec:polModel} and illustrated in Supplementary Figure \ref{fig:polModel}, comprises an elliptical retarded followed by a partial polariser. In the Jones formalism these are each represented by $2\times2$ matrices, which are multiplied together to obtain overall behaviour.  However, matrix multiplication is non-commutative and so it is necessary to consider an alternative model in which the order of these two elements is reversed.  By adapting Equation \ref{eq:jonesMat} we get an alternative composite Jones matrix:

\begin{equation}
\label{eq:jonesMat_rev}
\mathbf{\hat{J}}(x,y) = \mathbf{A}_{ret} \mathbf{A}_{pol} 
\end{equation}

If we then expand both Equation \ref{eq:jonesMat} and \ref{eq:jonesMat_rev} using Equations \ref{eq:polMat} and \ref{eq:retMat}, we get:

\begin{equation}
\label{eq:polOrder1}
\mathbf{J}(x,y) = \left(\begin{array}{c c}
K & L \\
M & N 
\end{array}\right)
\end{equation}

\noindent and

\begin{equation}
\mathbf{\hat{J}}(x,y) = \left(\begin{array}{c c}
\hat{K} & \hat{L} \\
\hat{M} & \hat{N} 
\end{array}\right)
\end{equation}

\noindent where

\begin{equation}
\begin{array}{c}
K = \left({\sin\left(\theta_D\right)}^2 \sqrt{1 - D} + {\cos\left(\theta_D\right)}^2 \sqrt{D + 1}\right) \left(\cos\left(\frac{\eta}{2}\right) + \cos\left(2 \theta_{\eta}\right) \sin\left(\frac{\eta}{2}\right) i\right) \\
- \frac{1}{2} \sin\left(\frac{\eta}{2}\right) \sin\left(2 \theta_{\eta}\right) \sin\left(2\theta_D\right) \left(\sqrt{D + 1} - \sqrt{1 - D}\right) \left(\sin\left(\xi\right) + i\cos\left(\xi\right)\right)
\end{array}
\end{equation}

\begin{equation}
\begin{array}{c}
\hat{K} = \left({\sin\left(\theta_D\right)}^2 \sqrt{1 - D} + {\cos\left(\theta_D\right)}^2 \sqrt{D + 1}\right) \left(\cos\left(\frac{\eta}{2}\right) + \cos\left(2 \theta_{\eta}\right) \sin\left(\frac{\eta}{2}\right) i\right) \\
- \frac{1}{2} \sin\left(\frac{\eta}{2}\right) \sin\left(2 \theta_{\eta}\right) \sin\left(2\theta_D\right) \left(\sqrt{D + 1} - \sqrt{1 - D}\right) \left(- \sin\left(\xi\right) + i\cos\left(\xi\right)\right)
\end{array}
\end{equation}

\begin{equation}
\begin{array}{c}
L = i \mathrm{e}^{i \xi} \sin\left(\frac{\eta}{2}\right) \sin\left(2\theta_{\eta}\right) \left({\sin\left(\theta_D\right)}^2 \sqrt{1 - D} + {\cos\left(\theta_D\right)}^2 \sqrt{D + 1}\right) \\
- \frac{1}{2} \sin\left(2\theta_D\right) \left(\sqrt{D + 1} - \sqrt{1 - D}\right) \left(\cos\left(\frac{\eta}{2}\right) - i\cos\left(2 \theta_{\eta}\right) \sin\left(\frac{\eta}{2}\right)\right)
\end{array}
\end{equation}

\begin{equation}
\begin{array}{c}
\hat{L} = i\mathrm{e}^{i\xi} \sin\left(\frac{\eta}{2}\right) \sin\left(2 \theta_{\eta}\right) \left({\sin\left(\theta_D\right)}^2 \sqrt{D + 1} + {\cos\left(\theta_D\right)}^2 \sqrt{1 - D}\right) \\
- \frac{1}{2}\sin\left(2 \theta_D\right) \left(\sqrt{D + 1} - \sqrt{1 - D}\right) \left(\cos\left(\frac{\eta}{2}\right) + i\cos\left(2 \theta_{\eta}\right) \sin\left(\frac{\eta}{2}\right)\right)
\end{array}
\end{equation}

\begin{equation}
\begin{array}{c}
M = \mathrm{e}^{i \xi} \sin\left(\frac{\eta}{2}\right) \sin\left(2 \theta_{\eta}\right) \left({\sin\left(\theta_D\right)}^2 \sqrt{D + 1} + {\cos\left(\theta_D\right)}^2 \sqrt{1 - D}\right) \\
- \cos\left(\theta_D\right) \sin\left(\theta_D\right) \left(\sqrt{D + 1} - \sqrt{1 - D}\right) \left(\cos\left(\frac{\eta}{2}\right) + i\cos\left(2 \theta_{\eta}\right) \sin\left(\frac{\eta}{2}\right)\right)
\end{array}
\end{equation}

\begin{equation}
\begin{array}{c}
\hat{M} = \mathrm{e}^{i \xi} \sin\left(\frac{\eta}{2}\right) \sin\left(2 \theta_{\eta}\right) \left({\sin\left(\theta_D\right)}^2 \sqrt{1 - D} + {\cos\left(\theta_D\right)}^2 \sqrt{D + 1}\right) \\
-  \cos\left(\theta_D\right)\sin\left(\theta_D\right) \left(\sqrt{D + 1} - \sqrt{1 - D}\right) \left(\cos\left(\frac{\eta}{2}\right) + i\cos\left(2 \theta_{\eta}\right) \sin\left(\frac{\eta}{2}\right)\right)
\end{array}
\end{equation}

\begin{equation}
\begin{array}{c}
N = \left({\sin\left(\theta_D\right)}^2 \sqrt{D + 1} + {\cos\left(\theta_D\right)}^2 \sqrt{1 - D}\right) \left(\cos\left(\frac{\eta}{2}\right) + i\cos\left(2 \theta_{\eta}\right) \sin\left(\frac{\eta}{2}\right)\right)\\
 -\frac{1}{2}\sin(\frac{\eta}{2})\sin\left(2\theta_D\right)  \sin\left(2\theta_{\eta}\right)\left( \sqrt{D + 1} - \sqrt{1 - D}\right) \left(-\sin(\xi)+i\cos(\xi) \right)
\end{array}
\end{equation}

\begin{equation}
\begin{array}{c}
\hat{N} = \left({\sin\left(\theta_D\right)}^2 \sqrt{D + 1} + {\cos\left(\theta_D\right)}^2 \sqrt{1 - D}\right) \left(\cos\left(\frac{\eta}{2}\right) + i\cos\left(2 \theta_{\eta}\right) \sin\left(\frac{\eta}{2}\right)\right)\\
 - \frac{1}{2}\sin\left(\frac{\eta}{2}\right) \sin\left(2 \theta_D\right) \sin\left(2 \theta_{\eta}\right) \left(\sqrt{D + 1} - \sqrt{1 - D}\right) \left(\sin\left(\xi\right) + i\cos\left(\xi\right)\right)
\end{array}
\end{equation}

Taking the difference between these two matrices, we find that:

\begin{equation}
\label{eq:polDiffMat}
\mathbf{J}(x,y) - \mathbf{\hat{J}}(x,y) = \mathbf{J'}(x,y) = \left(\begin{array}{c c}
K' & L' \\
M' & N' 
\end{array}\right)
\end{equation}

\noindent where

\begin{equation}
\begin{array}{c}
K' = K - \hat{K} = - \left(\sqrt{D + 1} - \sqrt{1 - D}\right) \sin\left(\frac{\eta}{2}\right) \sin\left(2 \theta_{\eta}\right) \sin\left(2\theta_D\right) \sin\left(\xi\right)
\end{array}
\end{equation}

\begin{equation}
\begin{array}{c}
L' = L - \hat{L} = \left(\sqrt{D + 1} - \sqrt{1 - D}\right)\left( i \mathrm{e}^{i \xi} \sin\left(\frac{\eta}{2}\right) \sin\left(2\theta_{\eta}\right)\cos(2\theta_D)+ i\sin\left(2\theta_D\right) \left(\cos\left(2 \theta_{\eta}\right) \sin\left(\frac{\eta}{2}\right)\right)\right)
\end{array}
\end{equation}

\begin{equation}
\begin{array}{c}
M' = M - \hat{M} = -\left(\sqrt{D + 1} - \sqrt{1 - D}\right) \mathrm{e}^{i \xi} \sin\left(\frac{\eta}{2}\right) \sin\left(2 \theta_{\eta}\right) \cos\left(2\theta_D\right) 
\end{array}
\end{equation}

\begin{equation}
\begin{array}{c}
N' = N - \hat{N} = \left( \sqrt{D + 1} - \sqrt{1 - D}\right) \sin(\frac{\eta}{2})\sin\left(2\theta_D\right)  \sin\left(2\theta_{\eta}\right) \sin(\xi)
\end{array}
\end{equation}

The error terms $K'$, $L'$, $M'$ and $N'$ are all proportional to $\left( \sqrt{D + 1} - \sqrt{1 - D}\right)$, and so as $D \to 0$, so these error terms become negligible.  To investigate the range of $D$ for which the error between these two models is negligible, we performed a Monte-Carlo simulation over the broadest possible range of parameters:

\begin{equation*}
\theta_{\eta} \sim U(-\pi/2,\pi/2)
\end{equation*}

\begin{equation*}
\eta \sim U(-\pi,\pi)
\end{equation*}

\begin{equation*}
\xi \sim U(-\pi,\pi)
\end{equation*}

\begin{equation*}
\theta_D \sim U(-\pi/2,\pi/2)
\end{equation*}

\begin{equation*}
D \sim N(0,\sigma_D)
\end{equation*}

100,000 samples were drawn from these distributions for different values of $\sigma_D$ using a MATLAB script.  For each sample drawn, Jones matrices are computed for the two orderings ($\mathbf{J}$ and $\mathbf{\hat{J}}$ respectively) of the polariser and retarder, enabling computation of the error matrix $\mathbf{J}'$ of Equation \ref{eq:polDiffMat}.  The elements are then normalised and summed giving an error metric, $Q$, defined (with reference to Equations \ref{eq:polOrder1} and \ref{eq:polDiffMat}) as:

\begin{equation*}
Q = \frac{1}{4} \left( \left|\frac{K'}{K}\right| + \left|\frac{L'}{L}\right| + \left|\frac{M'}{M}\right| + \left|\frac{N'}{N}\right| \right)
\end{equation*}

\begin{equation}
Q = \frac{1}{4} \left( \left|\frac{K - \hat{K}}{K}\right| + \left|\frac{L - \hat{L}}{L}\right| + \left|\frac{M - \hat{M}}{M}\right| + \left|\frac{N - \hat{N}}{N}\right| \right)
\end{equation}

This represents the average absolute proportional deviation introduced by reversing polarisation elements.  The measured cumulative distribution function for different values of $\sigma_D$ is shown in Supplementary Figure \ref{fig:polMonteCarlo}.  In the samples of mouse oesophageal tissue presented here, the true value of $\sigma_D$ is estimated to be 0.0081.  As this is less than 0.01, it is observed from the cumulative density function that in $>$99.5\% of possible cases the error introduced by a different ordering of polarisation elements will be $<$5\%. If a 10\% error threshold is considered acceptable, then a value of $\sigma_D = $0.05, 5-times the measured value, would still be sufficiently accurate in $>$96\% of cases.  Of course, if samples are expected to have larger values of $D$ then the two orderings would give different results and it would be necessary to determine which ordering had the stronger biological underpinning and best explained the measured data.  Such a `model comparison' could be achieved through an extension of the Bayesian inference approach of Supplementary Section \ref{subsec:bayesInf}.

\subsection{Bayesian inference}
\label{subsec:bayesInf}
The polarisation model is fitted to the measured data in order to extract the 5 polarisation properties ($\theta_D$, $D$, $\xi$, $\theta_{\eta}$, $\eta$) using a Bayesian inference approach, which has been found to be more robust to noise than clustering or pseudo-inverse approaches \cite{Zallat2008,Kasaragod2014}. In this work maximum likelihood estimates are determined during this inference and are used to plot images.  These could be computed more quickly by deriving analytical expressions via eigenvalue decomposition of the Jones matrix \cite{Makita2010, Lu1996}.  However, such an approach does not easily permit estimation of joint posterior probabilities of neighbouring pixels, which is essential for compensating beam misalignment (see Section \ref{subsubsec:beamMisalignment}). Therefore, we numerically infer joint maximum likelihood values.

To fit the model, we evaluate the \emph{posterior} probability of the parameter set as a function of the data using Bayes' theorem:

\begin{align*}
p\left[ D, \theta_D, \eta, \theta_{\eta}, \xi | \mathbf{U}(x,y), \mathbf{V}(x,y)\right. & \left.\right] \propto \\
p\left[\mathbf{V}(x,y) = \right. & \left. \mathbf{J}(D, \theta_D, \eta, \theta_{\eta}, \xi,x,y) \mathbf{U}(x,y)\right] \cdot p(D, \theta_D, \eta, \theta_{\eta}, \xi)
\end{align*}

The values in $\mathbf{V}(x,y)$ represent measured complex quantities and so are assumed to be independent and drawn from complex Gaussian distributions as:

\begin{equation}
\label{eq:complexNormal}
\mathbf{v}(x,y) \sim \mathcal{CN}\left[\mathbf{J}(D, \theta_D, \eta, \theta_{\eta}, \xi,x,y) \mathbf{u}(x,y), \sigma^2 \mathbf{I}\right]
\end{equation}

\noindent where $\mathbf{v}(x,y)$ is a column of $\mathbf{V}(x,y)$, $\mathbf{u}(x,y)$ is a column of $\mathbf{U}(x,y)$, $\sigma^2 \mathbf{I}$ is the covariance matrix, and $\mathcal{CN} (\mathbf{\mu},\mathbf{\Sigma})$ is a 2-D complex Gaussian distribution of mean $\mathbf{\mu}$ and covariance $\mathbf{\Sigma}$. The noise standard deviation, $\sigma$ is an additional parameter that can be inferred from the data.  However, after initial processing of several samples, the value was fixed at its inferred maximum likelihood value of 0.4 for subsequent calculations in order to increase speed.  We wish to use a broad prior to minimise any bias in the results.  The broadest possible prior would be a 5-dimensional uniform distribution over all the polarisation parameters (ignoring $\sigma$ by assuming it is now fixed).  In this case, the prior distributions for each individual parameter are independent so that we can write:

\begin{equation}
p(D, \theta_D, \eta, \theta_{\eta}, \xi) = p(D)p(\theta_D)p(\eta)p(\theta_{\eta})p(\xi)
\end{equation}

It is assumed that in the case of a slightly more restrictive prior, this independence assumption will still be valid to a good approximation and so it used for the rest of the analysis presented here.  Priors for parameters representing angles, i.e. circular quantities, are modelled with \emph{von Mises} distributions. The von Mises distribution is a continuous probability distribution defined on a circular domain of $[-\pi,\pi]$ that wraps around, giving a probability density function of $\frac{e^{\kappa \cos (x-\mu)}}{2 \pi I_0(\kappa)}$ where $\mu$ is the mean, $1/\kappa$ is analogous to variance, and $I_0(\cdots)$ is a modified Bessel function of order 0. For $\theta_D$, $\theta_{\eta}$ and $\eta$, the $\kappa$ is set to $\infty$ providing the broadest possible prior. For $\xi$, a restricted value of $\kappa = 1$ is used.  This is because when performing inference of known test targets, it was found that this parameter is prone to over-fitting and that a slightly restrictive prior substantially prevents this.  Further, previous studies that have shown that even in samples with high concentrations of chiral molecules such as glucose, the measured circularity is still relatively small \cite{Guo2007}.  It is therefore not expected that this parameter will be very large for the phantoms or tissue samples measured. Diattenuation, $D$, is the only non-circular parameter (i.e. it is not an angle) and so its prior is a truncated Gaussian over its domain $[-1,1]$ with mean 0 and broad variance.  Again, the broad (as opposed to infinite) variance is used to prevent overfitting and is consistent with previous work that has found the values of $D$ to be small (typically $<$0.2) in real tissue samples \cite{Qi2016,Pierangelo2013a}.  It is noted that the ability to find more accurate solutions in the presence of noise or incomplete information using such prior knowledge is a major benefit of the Bayesian inference approach.

The inference of parameters from data was performed numerically using the STAN package \cite{StanMan}, which produces samples from \emph{a posteriori} distributions using a Markov Chain Monte Carlo simulation engine.  It can also provide maximum likelihood estimates through an optimisation engine, which is used here on account of its high speed, and can also perform variational inference.  

\subsection{Compensation for beam misalignment}
\label{subsubsec:beamMisalignment}
The holographic endoscope design exploits two orthogonally polarised arms that are split then recombined before illuminating the sample. This is achieved using a custom housing for the mirrors and polarising beam splitters, which provides a compact way of transferring the beam path between two optical axes at different heights. As a result of this dual arm approach, any slight misalignment of the two polarisation arms leads to a spatially varying illumination polarisation state at the sample, illustrated in Supplementary Figure \ref{fig:waveplateImaging}a. For samples with small retardance this misalignment is not apparent because it is present for both the characterisation of the fibre transmission matrix and the sample imaging steps, and so the relative change in polarisation state is correct.

However, for samples that exhibit large values of retardance, $\eta$, and in which the retardance axes are not aligned with the polarisation axes of the illumination (i.e. ($\theta_{\eta} \neq 0,\pm\pi/2$), light is coupled between the slightly misaligned horizontally and vertically polarised beams. This produces, in the illumination polarisation basis, wavefronts with different tilts in the two polarisation arms.  This manifests as an apparent spatially varying retardance, illustrated in Supplementary Figure \ref{fig:waveplateImaging}a. This apparent retardance does not appear when the polarisation axes are aligned with the retardance axes. Therefore, the effect is removed by re-expressing Equation \ref{eq:forwardModelJones} in a linear polarisation basis at angle $\hat{\theta}_{\eta}$ to the illumination polarisation axes:

\begin{equation}
\label{eq:rotatedPolAxes}
\mathbf{V}(x,y) = \left. \mathbf{J}(x,y) \right|_{\theta_{\eta}=0}  
\left( \begin{array}{cc}
\cos \hat{\theta}_{\eta} & \sin \hat{\theta}_{\eta} \\
-\sin \hat{\theta}_{\eta} & \cos \hat{\theta}_{\eta} \end{array} \right)
\mathbf{U}(x,y)
\end{equation}


We then apply Bayesian inference in this new, rotated basis with $\hat{\theta}_{\eta}$ taking the place of $\theta_{\eta}$ as the inferred orientation angle of the retardance axes. To find the $\hat{\theta}_{\eta}$ that minimises the spatial tilt of the retardance (confirming the new basis is aligned with the retardance axes) we compute the joint probability for a small set of $R$ neighbouring pixels surrounding $(x,y)$.  For computational tractability, we employ a technique from variational Bayesian inference that approximates the posterior probability density function as a product of distributions \cite{MacKay2003}:

\begin{align*}
p[ D(x_1,y_1)\cdots D(x_R, y_R), &\theta_D(x_1,y_1)\cdots, \eta(x_1,y_1)\cdots, \hat{\theta}_{\eta}(x_1,y_1)\cdots, \xi(x_1,y_1)\cdots | \mathbf{U}(x_1,y_1)\cdots, \mathbf{V}(x_1,y_1)\cdots] \\
&= \prod_{r=1}^R p[D(x_r,y_r),\theta_D(x_r,y_r),\eta(x_r,y_r), \hat{\theta}_{\eta}(x_r,y_r), \xi(x_r,y_r) | \mathbf{U}(x_r,y_r), \mathbf{V}(x_r,y_r)]
\end{align*}

\noindent where the term inside the product is computed using equations \ref{eq:complexNormal} and \ref{eq:rotatedPolAxes}. This approximation implies independence of the conditional probabilities of polarisation parameters of neighbouring pixels.  This is a reasonable assumption because the conditional probability distributions (that is, the probability distributions given the experimentally measured Jones matrices) are effectively accounting for noise, which is assumed to be produced by an independent random process at each pixel.

The maximum likelihood probability is maximised when neighbouring pixels have similar conditional distributions for their polarisation parameters. This implicitly biases towards values of $\hat{\theta}_{\eta}$ where the spatial variation of retardance is minimised, i.e. when the means of the retardance distributions across neighbouring pixels are most similar.

It is not known \emph{a priori} whether the phantoms and tissue samples measured will exhibit significant birefringence.  However, the advantage of the above approach is that it will still provide the best estimate of parameters even if no birefringent tilt artefact is present. This is because the probability distribution of $\hat{\theta}_{\eta}$ tends towards being uniform when $\eta \xrightarrow{}$0, which is the same behaviour expected for $\theta_{\eta}$ (i.e. without the correction).  Therefore, the correction is applied to all measurements taken and will have more effect where it is naturally required.

The number of adjacent pixels used for the correction is set as $R$ = 15.  For optimal resolution, this value should be as small as possible because it effectively reduces resolution by low-pass filtering the image.  On the other hand, larger values can correct for smaller tilt artefacts.  $R$ = 15 was found experimentally to be the smallest value that was sensitive enough to correct for the degree of tilt artefact typically observed for the samples measured (which is in turn dependent on the degree of system misalignment). The outcome of this correction is validated by rotating a half-waveplate in the sample plane (Supplementary Figure \ref{fig:waveplateImaging}).

\subsection{Degeneracy of solutions}
We are decomposing a single matrix into the product of 8 matrices (Equations \ref{eq:jonesMat}, \ref{eq:polMat} and \ref{eq:retMat}), each comprising sinusoidal or complex exponential terms, hence there will invariably be degeneracy of the solutions. For example, consider the matrix $\mathbf{A}_{ret}$.  The 5 matrices of equation \ref{eq:retMat} can be multiplied to give:

\begin{equation}
\mathbf{A}_{ret} = \left( \begin{array}{cc}
e^{i\eta/2} \cos^2 \theta_{\eta} + e^{-i\eta/2} \sin^2 \theta_{\eta} & \left(e^{i\eta/2} - e^{-i\eta/2}\right) e^{-i \xi} \cos \theta_{\eta} \sin \theta_{\eta} \\
\left(e^{i\eta/2} - e^{-i\eta/2}\right) e^{i \xi} \cos \theta_{\eta} \sin \theta_{\eta} & e^{i\eta/2} \sin^2 \theta_{\eta} + e^{-i\eta/2} \cos^2 \theta_{\eta}  \end{array} \right)
\end{equation}

By substituting $\theta_{\eta} \rightarrow \theta_{\eta} + \pi/2$ and $\eta \rightarrow -\eta$, the first row, first column element of $\mathbf{A}_{ret}$ becomes:

\begin{equation*}
e^{-i\eta/2} \cos^2 (\theta_{\eta}+\pi/2) + e^{i\eta/2} \sin^2 (\theta_{\eta}+\pi/2)= e^{i\eta/2} \cos^2 \theta_{\eta} + e^{-i\eta/2} \sin^2 \theta_{\eta} 
\end{equation*}

The element of $\mathbf{A}_{ret}$ is therefore unchanged; the other 3 elements are also unchanged under this substitution.  Therefore, the parameter sets  $(\theta_{\eta}, \eta)$ and $(\theta_{\eta} + \pi/2, -\eta)$ are degenerate.  

Considering the ranges of all parameters, outlined in Equation \ref{eq:paramRanges}, there is in fact an 8-fold ambiguity for each point in the 5-D parameter space, comprising the following parameter sets:

\begin{equation}
\begin{array}{ccccccc}
(&\theta_{\eta}, &\eta,&\xi,&D,&\theta_D&)\\
(&\theta_{\eta}+\pi/2,&-\eta,&\xi,&D,&\theta_D&)\\
(&-\theta_{\eta},&\eta,&\xi+\pi,&D,&\theta_D&)\\
(&-\theta_{\eta}+\pi/2,&-\eta, &\xi+\pi,&D, &\theta_D&)\\
(&\theta_{\eta},&\eta, &\xi, &-D, &\theta_D + \pi/2&)\\
(&\theta_{\eta}+\pi/2, &-\eta, &\xi, &-D, &\theta_D + \pi/2&)\\
(&-\theta_{\eta}, &\eta, &\xi+\pi, &-D, &\theta_D + \pi/2&)\\
(&-\theta_{\eta}+\pi/2, &-\eta, &\xi+\pi, &-D, &\theta_D + \pi/2&)
\end{array}
\end{equation}

\noindent where parameters are defined to wrap around their ranges, listed in Equation \ref{eq:paramRanges}, via a modulus operation. If we were to randomly select one of the 8 degenerate solutions for each pixel, it could give the misleading impression that polarisation properties exhibit a high degree of variation.  To avoid this scenario, we select a point in 5-D parameter space (here we use $(\theta_{\eta} = \pi/4, \eta = \pi, \xi = 0, D = 1, \theta_D = \pi/4)$) and choose the degenerate solution closest to that point.  This ensures the maximum degree of smoothness in the resulting polarisation images. As a result, observed variations can be confidently classed as structural tissue features rather than artefacts. 

\pagebreak
\noindent
\Large\textbf{Supplementary Note 3:}\\
\Large\textbf{Preparation and characterisation of tissue mimicking phantoms}\\
\normalsize
\setcounter{section}{3}
\setcounter{subsection}{0}

\subsection{Preparation of optically scattering and birefringent phantoms}
All chemicals were purchased from Sigma Aldrich unless otherwise stated. Birefringent phantoms were made from acrylamide/bis-acrylamide (30\% concentration) solution as the base material. Ammonium persulfate (APS) and tetramethylethylenediamine (TEMED) were used as catalysts for the co-polymerisation of bis-acrylamide and acrylamide gels. Phantoms were fabricated inside a ventilated fumehood cupboard by mixing 30 mL of Acrylamide/Bis-acrylamide with 600 $\mu$L of APS and 40 $\mu$L of TEMED under thorough vortexing.  The mixture was poured immediately into petri dishes and left for approximately 30 minutes until the polymerisation was complete. Birefringent phantoms having thickness $\sim$2 mm were fabricated by pouring 5 mL of the mixture into the petri dishes. The set phantoms were then extracted from the petri dishes and were mounted on to the mechanical stretcher assembly as shown in Supplementary Figure \ref{fig:stretcher}.

Agar-based scattering phantoms were fabricated using 1.5 w/v\% agar solution as the base material as described elsewhere \cite{Joseph2017}. Briefly, agar powder was dissolved in de-ionised water and the solution was heated up to 100$^o$C inside a microwave oven. Pre-warmed intralipid and nigrosin dye solutions were added into the cooling agar solution to introduce scattering and absorption. A vortex mixer was used to homogenise the phantom solution and prevent separation of its components. The mixture was pipetted into petri dishes and the solutions were allowed to cool under room temperature to form phantoms having $\sim$2 mm thickness. 

\subsection{Characterisation of optical absorption and scattering}
A double integrating sphere (DIS) system\cite{Pickering1993} was used to determine the reduced scattering and absorption coefficients of the phantoms. The DIS set-up uses two highly reflective spheres (Labsphere), with Lambertian surfaces made from polytetrafluoroethylene (PTFE), to capture and quantify the amount of transmitted and reflected light. A broadband tungsten halogen lamp (AvaLight-HAL-MINI, Avantes) and two fiber-optic spectrometers (AvaSpec-ULS2048, Avantes) served as the light source and detectors respectively. The inverse-adding doubling algorithm (IAD) was used to compute the optical properties of the phantoms from the reflectance, transmittance and reference measurements \cite{Prahl1993}.

\subsection{Characterisation of stretching deformation and associated birefringence}
Deformation contours across the birefringent phantoms mounted in the stretcher were assessed by drawing a grid pattern onto the polyacrylamide gel using a marker pen. Qualitative observation of the contours after stretching revealed a macroscopic non-linearity in the deformation, however, over the small field of view of the endoscope the stretch was approximately linear. Due to the constant volume of the material, the change in the thickness upon stretching must be accounted for when plotting retardance, by considering the change in area between the two clamps and dividing the measured retardance by the inferred width.

\pagebreak
\noindent
\Large\textbf{Supplementary Note 4:}\\
\Large\textbf{Calculation of entropy and expectation in endoscopic images}\\
\normalsize
\setcounter{section}{4}
\setcounter{subsection}{0}
\subsection{Distribution fitting}
To identify lesions in phase and polarisation images, two image filters/operators are used (see main manuscript): spatial entropy, $H()$, and spatial expectation, $E()$.  These can both determined under a unified framework that estimates the probability distribution of spatial points and then provides the differential entropy of the inferred distribution as $H()$ and the mean as $E()$.  This framework is also easily extended to joint distributions, e.g. for co-occurrence metrics or for multivariate entropy.

The distribution fitting process is detailed in Supplementary Figure \ref{fig:entropyCalcAlg}.  The first step is to defined a spatial region (a set of $M$ pixels, $\mathbf{S}$) to which a distribution will be fitted. It must be decided whether or not the pixel values $\mathbf{S}$ are assumed to be statistically independent.  In the most general case, the $M$ pixels of $\mathbf{S}$, will have a joint distribution density function $p(\mathbf{z}_1, ..., \mathbf{z}_m, ..., \mathbf{z}_M)$, where $\mathbf{z}_m$ is a $P$-dimensional vector containing some subset of available quantities (amplitude, phase, retardance etc.).  Under this model, all $M$ pixels together effectively represent a single sample from an $M\cdot P$-dimensional distribution.  It is not typically feasible to fit such a distribution based on only a single sample and so some assumptions must be made to make the problem tractable.

A widely used assumption is that points separated by some fixed spatial vector $(a,b)^T$ are drawn from a joint distribution, $p(\mathbf{z}(x,y), \mathbf{z}(x+a,y+b))$.  This only has $2\cdot P$ dimensions and there are now $M$ samples for fitting (one for each point in $\mathbf{S}$). This assumption is regularly used in grayscale image amplitude analysis where it is termed the \emph{gray-level co-occurrence matrx (GLCM)} \cite{Albregtsen2008}.  In this work we therefore refer to quantities generated using this assumption as being \emph{co-occurrence} metrics.

A further relaxation from the co-occurrence case is to assume that each pixel, $\mathbf{z}$ in the area $\mathbf{S}$ is drawn independently from a distribution, $p(\mathbf{z})$. This, too, offers $M$ points for fitting.  However, as discussed in the main manuscript, it is found that there is relatively little difference between these two cases for the samples measured and so the latter is typically preferred.

If the quantitites chosen for fitting are all circular (i.e. represent angles), then a \emph{generalisd multivariate von Mises distribution} is used for fitting \cite{Mardia2008}. If only $A$ is chosen a chi distribution is fitted ($A$ is modelled as the absolute value of a Gaussian) and if only $D$ is chosen, a truncated Gaussian on the range $[-1,1]$ is fitted.  If a mixture of circular and non-circular quantities is used it is difficult in general to define a multivariate distribution with appropriate marginal distributions for all quantities. One solution is to scale the domain of the non-circular quantities such that their expected probability mass lies close to 0 on the $(-\pi,\pi]$ range.  In this case, as we let $\kappa \rightarrow 0$, the von Mises distribution starts to approximate a truncated Gaussian distribution which can be fit to the non-circular quantities.
 
Fitting is performed by evaluating predetermined analytic expressions for maximum likelihood parameters of distributions based on measured quantities. Such expressions are known for Gaussian and multivariate von Mises distributions \cite{Mardia2008}.

\subsection{Computation of entropy}
\label{subsec:entropyCalc}
Fitting distributions to points in a spatial region gives estimates for the mean and variance of a quantity in that region.  Given these, the \emph{entropy} of the distribution can then be computed, which gives a measure of spatial heterogeneity that can be indicative of disordered microstructure. There are many different definitions of entropy (e.g. Shannon entropy, thermodynamic entropy), although on some level they are fundamentally analogous. The definition used here is the \emph{differential entropy}, which is a property of the fitted distributions.  This metric can be derived from the Kullback-Leibler divergence, which measures the similarity of probability distributions $P$ and $Q$ with density functions $p(x)$ and $q(x)$ (the 1-D case) respectively:

\begin{equation}
\label{eq:KLdiv}
D_{KL}(P||Q) = \int_{-\infty}^{\infty} p(x) \log \frac{p(x)}{q(x)} dx
\end{equation}

If $Q$ is a uniform distribution then $D_{KL}(P||Q)$ represents how similar to a uniform distribution $P$ is -- that is, how 'spread out' $P$ is.  This is precisely what we are seeking: a measure of heterogeneity.  Setting $Q$ to be a uniform distribution and introducing a scale factor of -1, Equation \ref{eq:KLdiv} becomes the differential entropy, $H$:

\begin{equation}
\label{eq:shannonEnt}
H = - \int_{-\infty}^{\infty} p(x) \log p(x) dx
\end{equation}

This can be extended to multivariate distributions simply by integrating over the additional variables.  However, computation of this integral can be computationally expensive.  Fortunately, for many distributions analytical expressions for the entropy in terms of distribution parameters are known. This is true of the multivariate Gaussian and univariate von Mises distributions \cite{Mardia2008}, shown in Supplementary Figure \ref{fig:entropyCalcAlg}.  Where analytical expressions are not known (e.g. multivariate von Mises) Monte-Carlo sampling methods are used for efficient integral evaluation \cite{MacKay2003}

By computing entropy for a sliding window over the entire raw image, we produce an \emph{entropy filtered} image of the desired quantity. Entropy filtering has been successfully applied in the field of amplitude-only texture analysis for classification of diseased tissues \cite{Yogesan1996,DeArruda2013,Ganeshan2007}.

\subsection{Comparison of entropy and expectation}
If instead of plotting entropy as described in the previous section, we plot the mean of the fitted distributions at each spatial point, this gives images of the spatial expectation, $E$, effectively an averaging filtering on the same length scale as entropy. As shown in Supplementary Figure \ref{fig:CNRall} it was found experimentally that phase entropy is a good indicator of early tumourigenesis, whereas for polarisation properties, expectation offers better performance.

In the case of retardance, $\eta$, this could be because diseased tissue has a higher \emph{average} retardance due to, for example, increasingly dense collagen networks associated with early malignant transformation. The spatial expectation would then represent this change better than entropy.  Further, the anisotropic nature of polarising molecules such as collagen may mean they are preferentially aligned along some axis.  Even if this alignment is weak, we might then expect to see an increase in the average value of the retardance axis orientation, $\theta_{\eta}$ or diattenuation axis orientation, $\theta_D$.  This is consistent with the results of Supplementary Figure \ref{fig:CNRall} that show more contrast in the spatial expectation of $\theta_{\eta}$ and $\theta_{\eta}$ quantities than in the entropy.

\pagebreak
\noindent
\Large\textbf{Supplementary Note 5:}\\
\Large\textbf{Limitations of current design and routes to real-time application}\\
\normalsize
\setcounter{section}{5}
\setcounter{subsection}{0}

\subsection{Experimental imaging acquisition speed}
Currently, the acquisition time for an amplitude, phase and polarisation image is 8.3 s. This is because 11 images are required: 7 for establishing the phase of the horizontally polarised image and 4 for determining the phase difference between the horizontally and vertically polarised images. To infer polarimetric properties of a tissue sample, 3 such images are recorded (see Online Methods and Supplementary Note \ref{subsec:bayesInf}) taking a total of 24.9 s.  For real time operation, this would need to be reduced substantially to $<$0.2 s.  

The following modifications could be made to progress the operation towards real-time imaging. Using a camera with an automated high dynamic range imaging procedure, region of interest readout, and consequently higher frame rate could offer a >15 -fold increase in speed. Next, using an appropriately designed phase-mask in place of the current Fresnel lens, it would be possible to use only 2 images for phase retrieval and 3 for polarimetry, offering a further 2-fold speed up \cite{Horisaki2015,Zhang2007}.  Using only 2 illumination states instead of 3 offers a further reduction. Together these improvements would bring sample imaging time down to below the 0.2 s target. Further speed increases could be achieved through parallelisation of phase/polarisation retrieval by subdividing the camera sensor area, offering a trade-off with resolution \cite{Millerd2004}. At this point the switching speed of the liquid crystals in the SLM (60Hz frame rate) would become the limiting factor but this can be addressed by using high-speed (20kHz frame rate) digital micromirror devices (DMDs) \cite{Dremeau2015, Mitchell2016}.

\subsubsection{Transmission matrix recovery speed}
Although the fibre transmission matrix is typically stable for many months on an optical bench, in reality perturbations such as fibre bending would require periodic re-measurement of the transmission matrix. Our parallel spot characterisation system (described in Online Methods) already offers a speed-up of 12 over conventional spot or plane-wave scanning systems for recording fibre characterisation measurements. As opposed to methods used for MMFs, this approach scales to higher resolutions without requiring additional experimental measurements.

Currently, calibration measurements take 50.8 minutes but by making use of the optimisations discussed in the previous section this could be reduced to around 22 seconds. Re-characterisation may only need to be performed periodically at longer intervals, meaning it should not drastically affect total imaging time.  Further, parallelised compressed imaging techniques could be exploited with an array of high-speed photodiodes to move towards real-time characterisation if required \cite{Macfaden2016}.

\subsection{Image reconstruction speed}
To recover phase and polarisation information from the raw intensity images produced by the camera, a phase retrieval algorithm is used (see Online Methods). This currently takes around 67 s per single image (67 $\times$ 3 = 201 for full polarimetric imaging). The iterative algorithm used here (based on the Gerchberg-Saxton algorithm) requires around 200 iterations to converge. If we were to instead directly solve transport of intensity equations (TIE) this could be reduced to $<$0.1 s per single image \cite{Allen2001}. 

For transmission matrix recovery, it is necessary to solve 293 phase retrieval problems (81$\times$3 characterisation $+$ 50 phase stabilisation), which currently takes 5.5 hours. Again by using a TIE approach this could be reduced to 29.5 s.  Because these problems are highly parallelisable, this could be sped up another $\sim$100-fold by using state-of-the-art processors (e.g. Intel Xeon server with 8 CPUs with 24 cores each).  Further parallelisation could be achieved by using dedicated graphics processing units (GPUs). The actual recovery of the matrix from these images requires solving 81$\times$12$\times$2=1944 inverse problems -- one per row of the desired $\mathbf{A}^{-1}$ (Supplementary Note \ref{subsec:tMatRecovery}).  This currently takes 9.5 hours.  However, by downsampling the camera images (e.g. sampling one pixel for each fibrelet) and increasing the degree of parallelisaiton from the current factor of 8 to $>$100, it should be possible to reduce this to $\sim$30 s. This could be further reduced by adopting solution methods that directly reconstruct input fields as weighted sums of the calibration fields rather than inverting the full transmission matrix \cite{Choi2014}. 

Bayesian inference of sample polarimetric properties (described in Supplementary Note \ref{subsec:bayesInf}) is performed using the STAN package maximum likelihood optimiser, which requires 1.6 seconds to perform inference on a single pixel.  A full endoscope frame then takes 24 minutes to reconstruct.  Again, this problem is highly parallelisable and so could benefit from a 100-fold increase in speed on a state-of-the-art CPU. Further, by deriving analytical expressions for the joint maximum likelihood parameters of neighbouring pixels and using these instead of optimisation, inference time could be reduced to $<$0.2 s per image.

Computing the spatial entropy and expectation of an image, which involves fitting distributions as described in Supplementary Note \ref{subsec:entropyCalc}, currently takes 5 s per endoscope frame. Again, by utilising greater parallelisation this could be reduced to $<$0.1 s.
 
Finally, there are solution methods that do not require an explicit phase-retrieval step and can retrieve the desired transfer matrix based purely on intensity-only measurements. For the case where the matrix is known to comprise independent, complex Gaussian distributed entries a variational inference approach can be used \cite{Dremeau2015}. For the more generalised case in which the matrix is sparse, as it is here, algorithms such as \emph{generalised approximate message passing (GAMP)} may be applied directly to amplitude image sets \cite{Rangan2010, Schniter2015}. These could potentially reduce reconstruction time, as well as requiring minimal experimental measurements.

\subsection{Reflection mode operation}
For \emph{in vivo} clinical use, the endoscope must operate entirely in reflection mode rather than its current transmission mode.  In principle, the transmission mode illumination of samples can be simulated in reflection by using oblique back-scattering illumination \cite{Ford2012}. However, the imaging fibre will invariably experience random bending and temperature variations when in use and so the transmission matrix will need to be re-characterised without access to the distal facet.


One approach to reflection mode operation is to attach a calibration beacon to the distal facet, which has been demonstrated to work for creating a single focussed spot in confocal imaging but does not in the same form support wide-field imaging \cite{Farahi2013}. Speckle correlation imaging does not require any calibration of fibres, but relies on the assumption that images are of real, positive targets with finite support, which is not true in phase imaging where objects are generally complex \cite{Porat2016}. Another approach is to attach a known reflective plate to the distal facet and use back reflected light to make dynamic adjustments to a pre-recorded transmission matrix \cite{Gu2015}, however, this has not yet been demonstrated experimentally. Initial work using highly-spaced MCF has shown promise for two-photon imaging \cite{Warren2016}. Experimental investigation of these approaches is underway in our laboratories as a next step towards clinical translation.



\bibliographystyle{naturemag}
\end{document}